\documentclass[twocolumn,showpacs,aps,prd,superscriptaddress]{revtex4}

\usepackage{placeins}
\usepackage{graphicx}
\usepackage{dcolumn}
\usepackage{amsmath}
\usepackage{epsfig}
\usepackage{longtable}
\usepackage{float}
\usepackage{enumitem}

\input babarsym

\def\omeg{\ensuremath{\omega}}
\def\kaon{\ensuremath{K^{(*)\pm}}}
\def\charm{\ensuremath{D^{(*)}}}

\def\pom {\ensuremath{\pm}\xspace}
\def\Deltam {\ensuremath{\Delta m}\xspace}
\def\calF{{\ensuremath{\cal F}}\xspace}

\def\pimp {\ensuremath{\pi^\mp}\xspace}

\def\Thrust{{\rm Thrust}}

\newcommand{\kevcc}{\ensuremath{{\mathrm{\,ke\kern -0.1em V\!/}c^2}}\xspace}

\newcommand{\BABARPubYear} {08}
\newcommand{\BABARPubNumber} {017}

\newcommand{\SLACPubNumber} {13305}
\newcommand{\LANLNumber} {0807.2408v2}

\def\figurebox#1#2#3{%
    \def\arg{#3}%
    \ifx\arg\empty
    {\hfill\vbox{\hsize#2\hrule\hbox to #2{\vrule\hfill\vbox to #1{\hsize#2\vfill}\vrule}\hrule}\hfill}%
    \else
    {\hfill\epsfbox{#3}\hfill}%
    \fi}

\begin{document}
\preprint{\babar-PUB-\BABARPubYear/\BABARPubNumber}
\preprint{SLAC-PUB-\SLACPubNumber}
\preprint{arXiv:\LANLNumber}

\begin{flushleft}
 \babar-PUB-\BABARPubYear/\BABARPubNumber\\
 SLAC-PUB-\SLACPubNumber\\
\end{flushleft}

\title{{\large \textbf{Measurement of Ratios of Branching Fractions and
\boldmath{\CP}-Violating Asymmetries of \boldmath{$\Bpm \to
D^{*}\Kpm$} Decays}}}

%
\author{B.~Aubert}
\author{M.~Bona}
\author{Y.~Karyotakis}
\author{J.~P.~Lees}
\author{V.~Poireau}
\author{E.~Prencipe}
\author{X.~Prudent}
\author{V.~Tisserand}
\affiliation{Laboratoire de Physique des Particules, IN2P3/CNRS et Universit\'e de Savoie, F-74941 Annecy-Le-Vieux, France }
\author{J.~Garra~Tico}
\author{E.~Grauges}
\affiliation{Universitat de Barcelona, Facultat de Fisica, Departament ECM, E-08028 Barcelona, Spain }
% ...
\collaboration{The \babar\ Collaboration}
\noaffiliation

\date{\today}

\begin{abstract}
We report a study of $\Bpm \to D^{*} \Kpm$ decays with $D^{*}$
decaying to $D \pi^0$ or $D \gamma$, using $383 \times 10^{6} \B\Bbar$
pairs collected at the $\FourS$ resonance with the \babar\ detector at
the SLAC PEP-II \B-Factory.
The $D$ meson decays under study include a non-\CP mode ($K^\pm\pi^\mp$),
\CP-even modes ($K^\pm K^\mp$, $\pi^\pm \pi^\mp$) and \CP-odd modes
($\KS\pi^0$, $\KS\phi$, $\KS\omega$).
We measure ratios ($R^*_{\CP\pm}$) of branching fractions of decays to \CP eigenmode
states and to flavor-specific states as well as \CP
asymmetries ($A^*_{\CP\pm}$). These measurements are sensitive to the unitarity
triangle angle \g.
We obtain
$ A^*_{\CP+} = -0.11 \pm 0.09 \pm 0.01$,
$ R^*_{\CP+} = ~1.31 \pm 0.13 \pm 0.04$, and
$ A^*_{\CP-} = 0.06 \pm 0.10 \pm 0.02$,
$ R^*_{\CP-} = ~1.10 \pm 0.12 \pm 0.04$,
where the first error is statistical and the second error is systematic.
Translating our results into an alternative parametrization,
widely used for related measurements, we obtain
$x_{+}^*=0.11\pm 0.06 \pm 0.02$ and $x_{-}^* = 0.00 \pm 0.06 \pm 0.02$.
No significant \CP-violating charge asymmetry is found in either the
 flavor-specific mode $D\to\Kpm\pimp$ or in $\Bpm\to \Dstar\pipm$
 decays.
\end{abstract}


\maketitle

\section{Introduction \label{sec:intro}}

In the Standard Model (SM), \CP-violating phenomena are a consequence of a
single complex phase in the Cabibbo-Kobayashi-Maskawa (CKM)
quark-mixing matrix~\cite{ref:CKM}.
The $\Bpm\to\charm\kaon$ decay modes provide a theoretically clean
determination of the unitarity triangle angle \g, since the latter is equal to the relative phase between the
CKM- and color-favored $b\to c$ and the CKM- and color-suppressed
$b\to u$ decay amplitudes that are dominant in the considered
decays.
The method proposed by Gronau, London and Wyler (GLW) makes use of the
direct \CP violation in the interference between the amplitudes
for $\Bpm{\to} \Dz\Kpm$ and $\Bpm{\to} \Dzb \Kpm$ decays when the \Dz
and $\Dzb$ mesons decay to the same \CP eigenstate
\cite{Gronau:1990ra,Gronau:1991dp}.
The same approach is equally applicable when the $D$ and/or the $K$ meson
is replaced with its excited state. In this paper we use the $\Bpm\to\Dstar \Kpm$ decay.
We use the notation \Dz, \Dstarz, \Dzb and \Dstarzb to denote states with
definite flavor,
while $D_{\CP+}$ and $\Dstar_{\CP+}$ denote \CP-even eigenstates, $D_{\CP-}$ and $\Dstar_{\CP-}$
denote \CP-odd eigenstates, and
$D$ and \Dstar denote any state of the $D(1864)^0$ and $D^*(2007)^0$ mesons,
respectively.
With the integrated luminosity presently available, it is not possible
to make a precise \g measurement with the GLW method alone, but the
combination of several methods and of several modes allows an improvement of
the overall precision~\cite{Atwood:2003jb}.

In the case of $\Bpm\to \Dstar\Kpm$ decays, one defines
the \CP-violating charge asymmetry
\begin{equation} \small
A^*_{\CP\pm}\equiv \frac
{\BR(\Bm \to D^{*}_{\CP\pm} \Km)-\BR(\Bp \to D^{*}_{\CP\pm}\Kp)}
{\BR(\Bm \to D^{*}_{\CP\pm} \Km)+\BR(\Bp \to D^{*}_{\CP\pm}\Kp)},
\label{eq:def:AstarCP}
\end{equation}
and the ratio of branching fractions for the decays to \CP
eigenmodes and flavor-specific states
\begin{equation} \small
R^{*}_{\CP\pm} \equiv \frac
{\BR(\Bm \to D^{*}_{\CP\pm} \Km) +\BR(\Bp \to D^{*}_{\CP\pm} \Kp) }
{\left[\BR(\Bm \to D^{*0} \Km) +\BR(\Bp \to \Dstarzb \Kp)\right]/2 }.
\label{eq:def:Rstar:pm}
\end{equation}
We refer to the companion of the charmed meson in the final state as
the prompt track.
Experimentally, it is convenient to normalize the branching fractions
 of the decays with a prompt kaon in the final state to those of the
 similar decays with a prompt pion that have a larger branching
 fraction.
The ratio $R^{*}_{\CP\pm}$ can then be expressed as
\begin{equation}
R^{*}_{\CP\pm}
\approx \frac{R^*_{\pm}} {R^*},
\label{eq:def:Rstar:pm:rap}
\end{equation}
where $R^*_{\pm}$ and $R^*$ are the $K/\pi$ ratios
\begin{equation}
R^*_{\pm} \equiv \frac
{\BR(\Bm \to D^{*}_{\CP\pm} \Km)+\BR(\Bp \to D^{*}_{\CP\pm} \Kp)}
{\BR(\Bm \to D^{*}_{\CP\pm}\pim)+\BR(\Bp \to D^{*}_{\CP\pm}\pip)},
\label{eq:def:RstarCP}
\end{equation}
and
\begin{equation}
R^* \equiv \frac
{\BR(\Bm \to D^{*0} \Km)+\BR(\Bp \to \Dstarzb \Kp)}
{\BR(\Bm \to D^{*0}\pim)+\BR(\Bp \to \Dstarzb\pip)}.
\label{eq:def:Rstar}
\end{equation}
The ratio $R^*$ is predicted to be of the order of
$\left[(f_{K}/f_{\pi})\times |V_{us}/V_{ud}|\right]^2 = 0.080 \pm 0.002$~\cite{ref:pdg06},
where $f_{K}$ and $f_{\pi}$ are the form factors of the mesons.
Equation (\ref{eq:def:Rstar:pm:rap}) would be an equality if \CP
violation was completely absent in $B\to \Dstar\pi$ decays.
Defining the charge asymmetry
\begin{equation} \small
A_h^*\equiv \frac
{\BR(\Bm \to D^{*} h^-)-\BR(\Bp \to D^{*}h^+)}
{\BR(\Bm \to D^{*} h^-)+\BR(\Bp \to D^{*}h^+)},
\label{eq:def:Astar}
\end{equation}
(noted $A^*_{\pi}$ and $A^*_{K}$ when referring to $h=\pi$ and $h=K$ respectively),
this approximation implies that the pion charge asymmetry $A^*_{\pi}$
should be compatible with zero, as should be the kaon charge asymmetry $A^*_{K}$
for the flavor-specific modes $D\to \Kpm\pimp$.
The possible bias induced by this approximation is expected to be
small since the ratio of the amplitudes of the $\Bm \to \Dstarzb \pim$
and $\Bm \to \Dstarz \pim$ processes is predicted to be of the order
of 1\%~\cite{Gronau:2002mu} in the SM, and will be accounted for in the
systematic uncertainties.

Most experimental systematic uncertainties, such as
those related to the reconstruction of the \Dstar, and the uncertainties
on the $D$ decay branching fractions, cancel in the $K/\pi$ ratios $R^*$ and $R^*_{\pm}$.
\begin{table*}
\caption{\label{tab:previous:results}
Past measurements of parameters related to the measurement of \g in
$\Bpm\to\Dstar\Kpm$ decays by the GLW method. }
\begin{center}
\begin{ruledtabular}
\begin{tabular}{lcccccccccc}
 & $A^*_{\CP+}$ & $A^*_{\CP-}$ & $R^*_{\CP+}$ & $R^*_{\CP-}$ & $R^*$ \\
\hline
\noalign{\vskip1pt}
\babar\ \cite{Aubert:2004hu} & $-0.10\pm 0.23\,^{+0.03}_{-0.04}$ & -- & $1.06\pm 0.26\,^{+0.10}_{-0.09}$
 & -- & $0.0813 \pm 0.0040\,^{+0.0042}_{-0.0031} $\\
Belle \cite{Abe:2001waa,Abe:2006hc} & $-0.20 \pm 0.22 \pm 0.04 $ & $ 0.13 \pm 0.30 \pm 0.08 $ & $ 1.41 \pm 0.25 \pm 0.06 $ & $ 1.15 \pm 0.31 \pm 0.12 $ & 0.078 \pom 0.019 \pom 0.009
\end{tabular}
\end{ruledtabular}
\end{center}
\end{table*}
By neglecting the small~\cite{Aubert:2007wf,Staric:2007dt} \Dz -- \Dzb
mixing~\cite{Grossman:2005rp} and \CP violation in $D^0$ decays, $R^*_{\CP\pm}$ and
$A^*_{\CP\pm}$ are related to \g through
\begin{equation}
R^{*}_{\CP\pm} = 1 + r^{*2}_B \pm 2 r^*_B \cos\delta^*_B\cos\gamma,
\label{eq:rel:Rstar:pm:gamma}
\end{equation}
and
\begin{equation}
A^*_{\CP\pm} =
\frac{\pm 2 r^*_B \sin\delta^*_B\sin\gamma}
{1 + r^{*2}_B \pm 2 r^*_B \cos\delta^*_B\cos\gamma},
\label{eq:rel:AstarCP:gamma}
\end{equation}
where $r^*_B$ is the magnitude of the ratio of the amplitudes for the
processes $\Bm\to \Dstarzb \Km$ and $\Bm\to \Dstarz \Km$, and
$\delta^*_B$ is the relative strong phase between these two
amplitudes.
The ratio $r^*_B$ involves a CKM factor
$|V_{ub}V_{cs}/V_{cb}V_{us}|\approx 0.44\pm 0.05$~\cite{ref:pdg06} and a color
suppression factor that has been estimated to lie between
$0.26\pm 0.07\pm 0.05$~\cite{Browder:1996af} and 0.44
\cite{Gronau:2002mu}, so that $r^*_B$ is predicted to be in the range
$0.1 - 0.2$.
More recent calculations that take into account final state interactions
\cite{Chua:2005dt}
yield predictions of $r^*_B =0.09 \pom 0.02$.

The latest results by \babar\ and Belle are reported in references
\cite{Aubert:2004hu} and  \cite{Abe:2001waa,Abe:2006hc} respectively.
\babar\ used $123 \times 10^6~\BB$ pairs with $\Dstar\to D \piz$ and
 $D$ reconstructed in the \CP-even modes $\Kp\Km$ and $\pip\pim$, and
 non-\CP modes $\Kpm\pimp$, $\Kpm\pim \pip\pimp$ and $\Kpm\pimp\piz$.
Belle used $275 \times 10^6~\BB$ pairs with $\Dstar\to D \piz$ and $D$
reconstructed in the \CP-even modes $\Kp\Km$ and $\pip\pim$, \CP-odd
modes $\KS\piz$, $\KS\omeg$, $\KS\mphi$ and non-\CP modes $\Kpm\pimp$
\cite{Abe:2006hc}.
The results are summarized in Table \ref{tab:previous:results}.
Similar studies have been performed on the channels $\Bpm\to D \Kpm$
\cite{Aubert:2005rw,Abe:2006hc,Aubert:2008yk} and $\Bpm\to D \Kstarpm$ \cite{Aubert:2005cc}.

The \babar~ \cite{Aubert:2006am} and Belle~\cite{Collaboration:2008wy}
experiments have recently obtained estimates of $r^*_B$ and $\delta_B^*$
parameters from the overlap of the \Dz and \Dzb decays
in the Dalitz planes of some three-body $D$ decays.
\babar\ obtains $r_B^*=0.135 \pm 0.051$ and $\delta_B^* = (-63^{+28}_{-30})^\circ$,
while Belle obtains $r_B^* = 0.21\pm 0.08 \pm 0.02 \pm 0.05$ and
$\delta_B^* = (342^{+21}_{-23} \pm 4 \pm 23)^\circ$
(where the first error is statistical, the second is the experimental
systematic uncertainty and the third reflects the uncertainty on the $D$
decay Dalitz models).

In this paper, by using $(383 \pm 4)
\times 10^6~\BB$ pairs, we update the results of our previous study of
 $\Bpm{\to} \Dstar\Kpm$ decays \cite{Aubert:2004hu}  for $D$ decays to the \CP-even modes $\Kp\Km$, $\pip\pim$ and to the flavor-specific modes $\Kpm\pimp$, and
 we extend it to the \CP-odd modes $\KS\piz$, $\KS\omeg$ and $\KS\mphi$,
 and to $\Dstar\to D \g$.
Due to parity and angular-momentum conservation,
the \CP eigenvalue of the \Dstar is inferred from that of the $D$, when the
\CP eigenvalue of the neutral companion (\g or \piz) is taken into
account \cite{Bondar:2004bi}:
$\CP(\Dstar)=\CP(D)$ when $\Dstar\to D\piz$, and $\CP(\Dstar)=-\CP(D)$
when $\Dstar\to D\g$.

Experimental results can also be presented using the ``cartesian coordinates''
\begin{equation}
(x^*_\pm, y^*_\pm) \equiv
(r^*_B \cos(\delta^*_B\pm \gamma),r^*_B \sin(\delta^*_B\pm \gamma)),
\label{eq:def:x:y}
\end{equation}
which have the advantage of having Gaussian uncertainties, and of
being uncorrelated and unbiased ($r^*_B$, being positive, is biased
towards larger values in low precision measurements, whereas $x^*_\pm$
and $y^*_\pm$ show no such bias)~\cite{Aubert:2005iz}.
The parameters $x^*_\pm$ can be obtained from $R^*_{\CP\pm}$ and
 $A^*_{\CP\pm}$,
\begin{equation}
x^*_{\pm} = \frac
{R^*_{\CP+}(1\mp A^*_{\CP+}) - R^*_{\CP-}(1\mp A^*_{\CP-})}{4}.
\label{eq:x:gamma}
\end{equation}
The measurements presented in this paper have no direct sensitivity to $y^*_\pm$, in
contrast to Dalitz analyses. However an indirect constraint can be
obtained using
\begin{equation}
(r^*_B)^2 = x_{\pm}^{*2} + y_{\pm}^{*2}=
\frac{R^*_{\CP+}+ R^*_{\CP-} -2}{2}.
\label{eq:x:y:gamma}
\end{equation}
Note that there are four observables in these parameterizations, either
($A^*_{\CP+}$, $R^*_{\CP+}$, $A^*_{\CP-}$ and $R^*_{\CP-}$) or
($x^*_{+}$, $y^*_{+}$, $x^*_{-}$ and $y^*_{-}$), while there are only
three independent fundamental parameters (\g, $r^*_B$ and
 $\delta^*_B$).
The set of parameters must therefore fulfill one constraint, which can
be $\kappa =0$, where
\begin{equation}
\kappa \equiv R^*_{\CP+}A^*_{\CP+} + A^*_{\CP-}R^*_{\CP-}.
\label{eq:def:kappa}
\end{equation}

\section{The Dataset and \babar\ Detector\label{sec:dataset}}

The results presented in this paper are based on data collected
with the \babar\ detector at the \pep2\ asymmetric-energy \epem\ storage ring
 of the Stanford Linear Accelerator Center. At \pep2,
9.0 \gev\ electrons and 3.1 \gev\ positrons collide at a center-of-mass
energy of 10.58 \gev, which corresponds to the mass of the \FourS\ resonance.
The asymmetric beam energies result in a boost from the laboratory to the
center-of-mass frame of $\beta\gamma\approx 0.56$.
The dataset analyzed in this paper corresponds to an integrated luminosity of 347 \invfb
at the \FourS\ resonance.

The \babar\ detector is described in detail elsewhere~\cite{babarnim}.
Surrounding the interaction point is a five-layer double-sided silicon
vertex tracker (SVT), which measures the trajectories of charged
particles. A 40-layer drift chamber (DCH) provides measurements of the
momenta of charged particles.
Both the SVT and DCH are located inside a 1.5 T magnetic field provided by a solenoid magnet.
Charged hadron identification is achieved through measurements of
particle energy-loss in the tracking system and the Cherenkov angle
obtained from a detector of internally reflected Cherenkov light (DIRC).
A CsI(Tl) electromagnetic calorimeter (EMC) provides photon detection,
electron identification, and \piz reconstruction.
Finally, the instrumented flux return (IFR) of the magnet enables
discrimination of muons from pions.
For the most recent 134 \invfb of data, a fraction of the resistive
plate chambers constituting the muon system has been replaced by
limited streamer tubes~\cite{Andreotti:2005uq}.

We use Monte Carlo (MC) simulation to study the detector acceptance
and backgrounds. The MC uses the EvtGen generator~\cite{Evtgen} and
GEANT4~\cite{Agostinelli:2002hh} to simulate the passage of particles
through matter.

\section{Reconstruction of \B Candidates \label{sec:reco}}

We perform an exclusive reconstruction of the full \B meson decay chain, in
the modes described in the introduction, starting from the final
stable products (charged-particle tracks and neutral electromagnetic
deposits in the EMC).

The \piz candidates used to form an $\omega$, a $D$ or a \Dstar candidate are
reconstructed from pairs of photons with energies larger than $30 \mev$,
and shower shapes consistent with electromagnetic showers, with
invariant mass in the range $115 < m_{\g\g} < 150 \mevcc$.
In addition, the \piz candidates used to form a \Dstar
candidate are required to have center-of-mass frame momenta $p_{\g\g}^* < 450 \mevc$.
The $\omega$ candidates are reconstructed in the three-body decay
$\omega \to \pi^+\pi^-\pi^0$, with an invariant mass within $50 \mevcc$
of the world average \cite{ref:pdg06}.
We reconstruct $\KS \to \pip\pim$ candidates from pairs of oppositely
charged tracks that are consistent with having originated from a
common vertex position
and with an invariant mass within $25 \mevcc$ of the world average
\cite{ref:pdg06}.
We reconstruct $\phi\to \Kp\Km$ candidates from pairs of oppositely
charged tracks with particle identification inconsistent with a pion
hypothesis, that are consistent with having originated from a common
vertex position, and that have invariant mass within $30 \mevcc$ of
the world average \cite{ref:pdg06}.

Only two-body $D$ decays are considered in this study. The $D$
candidates are reconstructed from their two daughters that
are required to be consistent with having originated from a common
vertex position.
In the case of $D\to\KS\piz$, in which vertexing of the $\KS\piz$
system would yield a poor geometrical constraint, a beam spot
constraint is added to the fit in order to force the $D$ daughters to
originate from the interaction region.

The \Dstar candidates are formed from $D$ and \piz or \g candidates.
These photon candidates are required to have energies larger than $100
\mev$ and shower shapes consistent with electromagnetic showers.
The \Dstar candidates are required to fulfill
$130 < \Delta m < 170 \mevcc$ and $80 < \Delta m < 180 \mevcc$,
 respectively, where \Deltam is the invariant mass difference between
the \Dstar and the $D$ candidate.

The \piz, \KS, $D$ and \Dstar candidates are refitted with mass
constraints before their four-momenta are used to reconstruct the \B
decay chain.
We form \B candidates from \Dstar candidates and charged tracks,
 fitted with a beam spot constraint.
We characterize \B candidates by two kinematic variables: the
difference between the reconstructed energy of the \B candidate and
the beam energy in the center-of-mass frame
$\DeltaE_K \equiv E_B^*-\sqrt{s}/2$, and the beam-energy substituted mass
$\mes \equiv\sqrt{(s/2+{\bf p}_0\cdot {\bf p}_B)^2/E_0^2-{\bf p}_B^{2}}$,
where
($E_0$, ${\bf p_0}$) and ($E_B$, ${\bf p_B}$) are the four-momenta of the
$\Upsilon(4S)$ and \B meson candidate, respectively, the asterisk
denotes the $\Upsilon(4S)$ rest frame, and $\sqrt{s}$ is the
total energy in the $\Upsilon(4S)$ rest frame.
The subscript $K$ in $\DeltaE_K$ indicates that the kaon hypothesis
has been assumed for the prompt track in the computation of \DeltaE.
For a correctly reconstructed \B meson having decayed to a $\Dstar K$
final state, $\DeltaE_K$ is expected to peak near zero, with an
R.M.S. of about $16 \mev$, and \mes is expected to peak near the
$\B$-meson mass $5.279 \gevcc$, with an R.M.S. that is almost
independent of the channel and close to 3 \mevcc.
For a $\B\to \Dstar \pi$ decay reconstructed as $\B\to \Dstar K$ with
a correctly identified \Dstar, the $\DeltaE_K$ peak is shifted by
approximately $+50\mev$.
At reconstruction level, the loose requirements $5.2<\mes<5.3\gevcc$
and $|\DeltaE_K|< 0.2 \gev$ are applied to the $B$ meson candidate.

We form a Fisher discriminant \calF \cite{fisher} to distinguish
 signal events from the significant background due to $\epem \to
 \qqbar$ $(q = u,d,s,c)$ continuum events.
Six variables are used:
 \begin{itemize}
 \item
$L_0$ and $L_2$, the zeroth and second angular moments of the energy
flow around the \B thrust axis. They are defined as
 $\sum_i p_i$ and $\sum_i p_i \cos^2{\theta_i}$ respectively,
where $p_{i}$ is the momentum and $\theta_{i}$ is the angle with
respect to the thrust axis of the \B candidate, both in the center-of-mass frame,
for all tracks and neutral clusters not used to reconstruct the \B
meson;
 \item
$R_2$, the ratio of the second to the zeroth Fox-Wolfram moment
\cite{FoxWolfram} of charged tracks and neutral clusters in the center-of-mass frame;
 \item
 $|\cos{\theta_B}|$, where $\theta_B$ is the angle between the
 momentum of the \B candidate and the boost direction of the \epem
 center-of-mass frame;
 \item
$|\cos{\theta_{\Thrust}}|$, where $\theta_{\Thrust}$ is the angle
between the \B candidate thrust vector and the beam axis in the
center-of-mass frame;
 \item
$|\cos{\theta_{\rm T}}|$, where $\theta_{\rm T}$ is the angle between
the \B candidate thrust axis and the thrust axis of the rest of the
event in the center-of-mass frame (where the rest of the event corresponds
to reconstructed particles not associated with the $B$ candidate).
 \end{itemize}

\section{Selection of \B Candidates \label{sec:selec}}

After the preliminary event reconstruction, a large amount of
background remains in the signal candidate sample.
In this section we describe the additional selection criteria  used to reduce the background.

The selection of each $B\to\Dstar K$ decay mode is optimized
separately, by the maximization of the sensitivity
$S/ \sqrt{S+B+1}$, where $\sqrt{S+B+1}$ is a symmetrized approximation
of the Poisson uncertainty on the measurement of $S+B$.
The numbers $S$ and $B$ of signal and background expected events are
estimated from, respectively, high-statistics exclusive MC samples,
and a cocktail of generic \BpBm (with signal events removed), \BzBzb and
\qqbar MC samples.

In the optimization procedure, we include requirements on all variables,
including those that will be relaxed during the fit, and including
tightening requirements that have been made in the reconstruction stage.
Our optimization procedure is similar to that used in
Ref.~\cite{Aubert:2001pe}, and allows us to determine the optimal set
of variables as well as the optimal requirements on those
variables, by the examination of the signal-to-background ratio
distributions~\cite{Neyman:Pearson}.
The final set of variables on which we apply selection optimization is:
\begin{itemize}
\item
the \B candidate-related variables \mes and $\DeltaE_K$ introduced
above;
\item
the mass $m_{D^0}$ of the $D$ candidate before the mass constraint is
applied and the mass difference \Deltam;
\item
likelihood ratios for the prompt track, that are evaluated making use
of the Cherenkov angle information from the DIRC, and of the
\dedx information provided by the tracking system.
Explicitly, we compute likelihoods $\calL_h$ for particle
identification (PID) hypotheses $h=K,\pi,p$ and make requirements on
the ratios $\calL_K/(\calL_K+\calL_\pi)$ and
$\calL_K/(\calL_K+\calL_p)$. We also require that the track is not identified as an electron or muon;
\item
likelihood ratios for pion and kaon candidates that are daughters of two-body $D$ decays;
\item
the value of the Fisher variable \calF;
\item
the invariant masses of the \KS, \mphi, \piz and \omeg\ candidates,
 when relevant, and before the mass
constraints.
Furthermore, for decays involving \KS candidates, we include the ratio
of the flight length of \KS candidates in the transverse plane divided
by its uncertainty, and  require it to be larger
than two.
For decays involving $\omega$ candidates, we include
$|\cos(\theta_{\omega})|$, where $\theta_{\omega}$ is the angle
between the normal to the pion decay plane and the $D$ direction in
the $\omega$ rest-frame.
\end{itemize}
The selection requirements applied to these variables are
mode-dependent, except for the prompt track PID requirements
$\calL_K/(\calL_K+\calL_\pi)>0.9$ and $\calL_K/(\calL_K+\calL_p)>0.2$ that are applied to all $B \to D^* K$ channels.\\
The selection of the $\Bpm\to\Dstar \pipm$ modes is identical to that
of the $\Bpm\to\Dstar \Kpm$ modes, except for
the prompt-track PID that is reversed ($\calL_K/(\calL_K+\calL_\pi)<0.2$).

A fraction of the events have several \B candidates
selected: the average multiplicity varies from 1.07 to 1.66 for
$\Dstar \to D \piz$ and from 1.00 to 1.25 for $\Dstar\to D \g$,
depending on the channel.
We select the \B candidate that has the \B vertex fit
with the largest probability.
This best-candidate procedure is used during the optimization of the
selection that we have described above.
The probability of selection of the well-reconstructed signal candidate is mode dependent
and is in the range 56--72\% for $\Dstar \to D \piz$ decays and in
the range 68--81\% for $\Dstar\to D \g$ decays.

\section{Maximum likelihood fit \label{sec:analysis}}

The dominant contribution to the remaining background after event
selection is from \B decays, including a significant amount of
feed-across from $\Bpm\to\Dstar\pipm$ decays.
Therefore the measurement is performed with an unbinned likelihood
fit~\cite{Verkerke:2003ir,minuit} based on two variables that best
discriminate this background, namely $\DeltaE_K$ and a PID variable
$\mathcal{T_R}$ defined below.

As the prompt track PID likelihood ratio
$\mathcal R\equiv \calL_K/(\calL_K+\calL_\pi)$
is very strongly peaked near zero for pions and near one for
kaons, we use a pseudo-logarithmic change of variable
\begin{equation}
\label{Eq:ydef}
\mathcal{T_R}=\log_{10}\left(\frac{\mathcal R+\epsilon}{1-\mathcal R+\epsilon}\right).
\end{equation}
We include a small positive number $\epsilon=10^{-7}$, so that $\mathcal{T_R}$ is
always defined, with $\mathcal{T_R}=+7$ for $\mathcal R= 1$ (``perfect kaons'')
and $\mathcal{T_R}=-7$ for $\mathcal R= 0$ (``perfect pions'').

\begin{figure*}
\begin{center}
\includegraphics[width=0.22\linewidth]{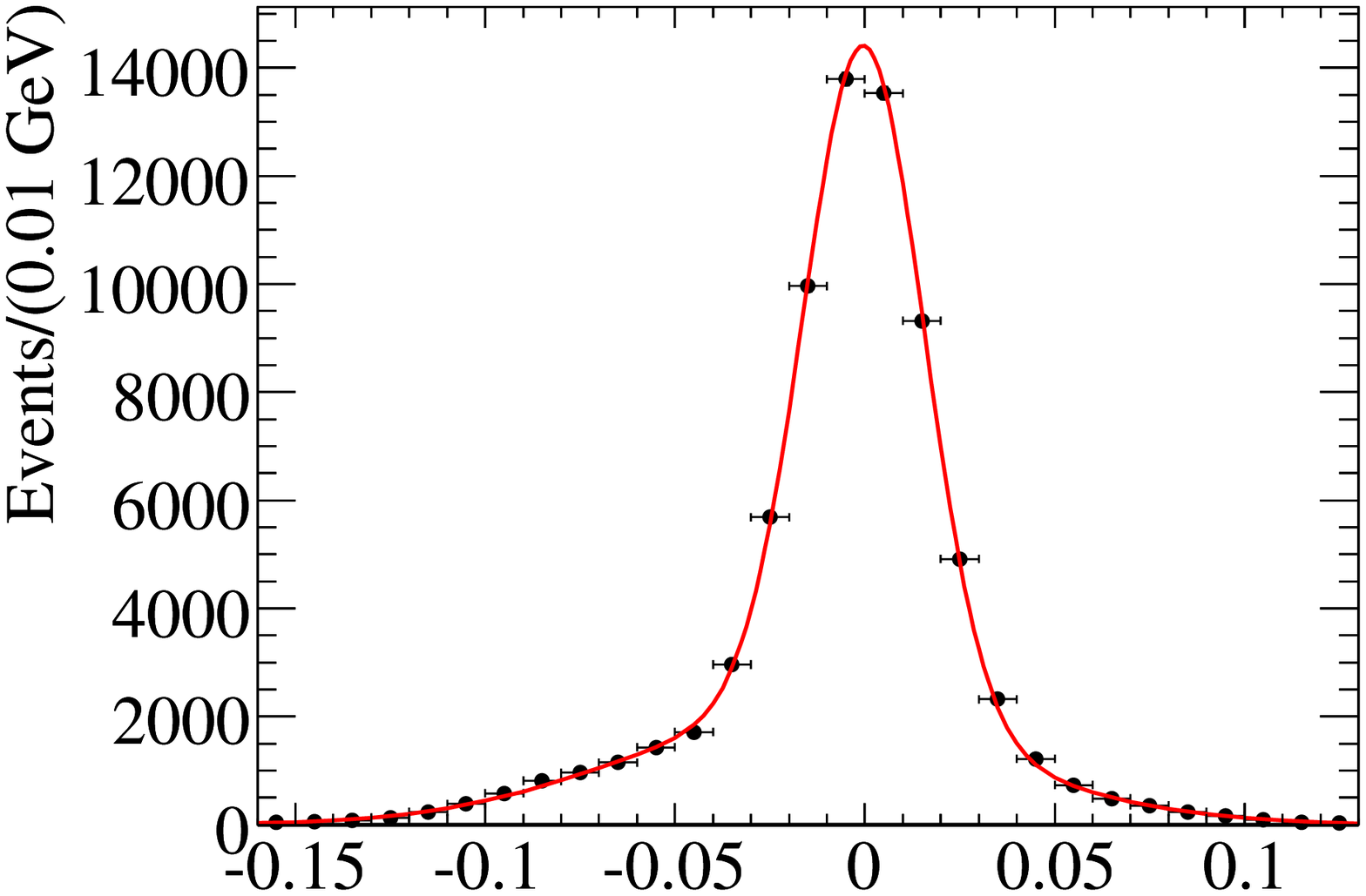}
\includegraphics[width=0.22\linewidth]{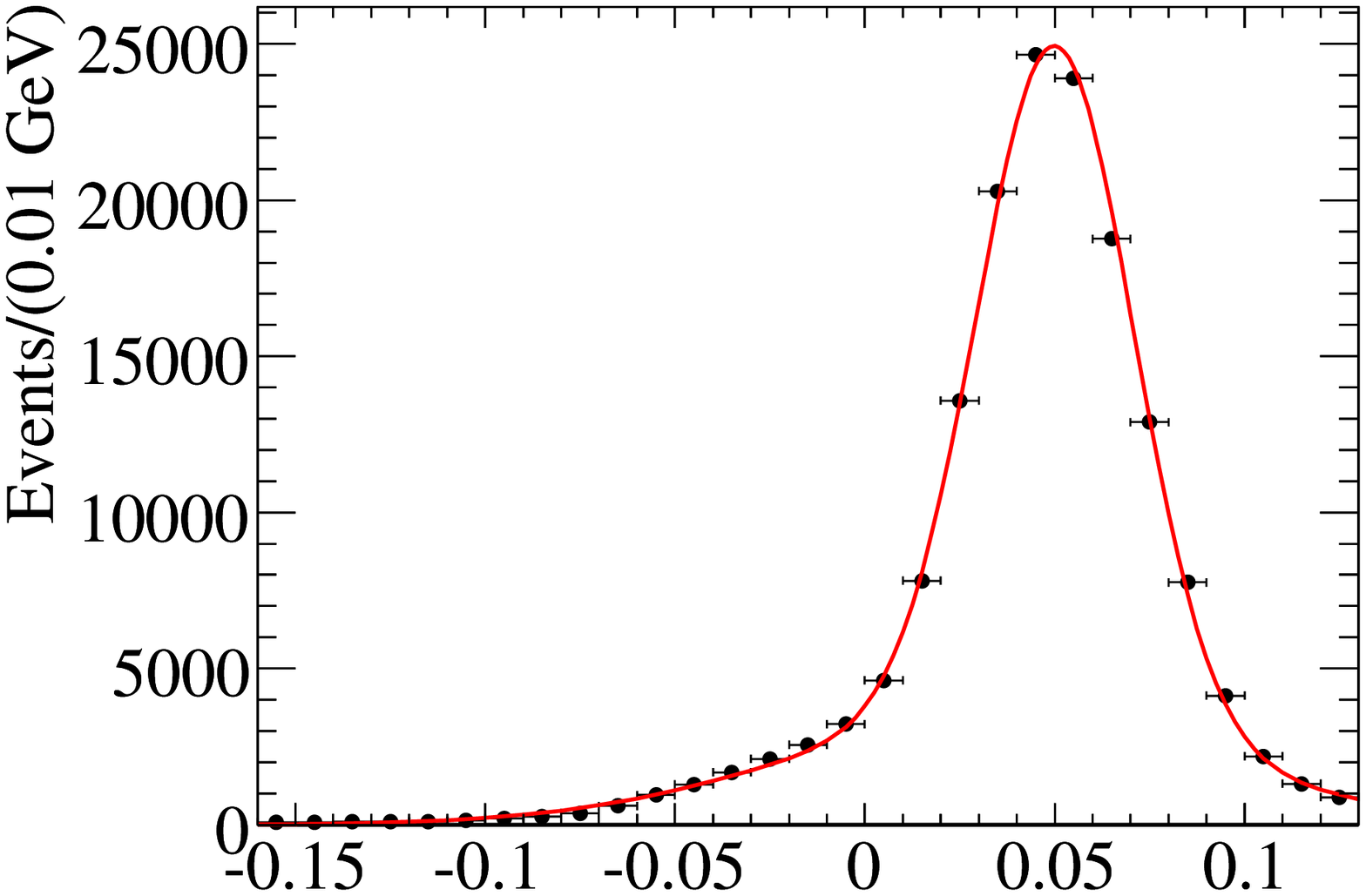}
\includegraphics[width=0.22\linewidth]{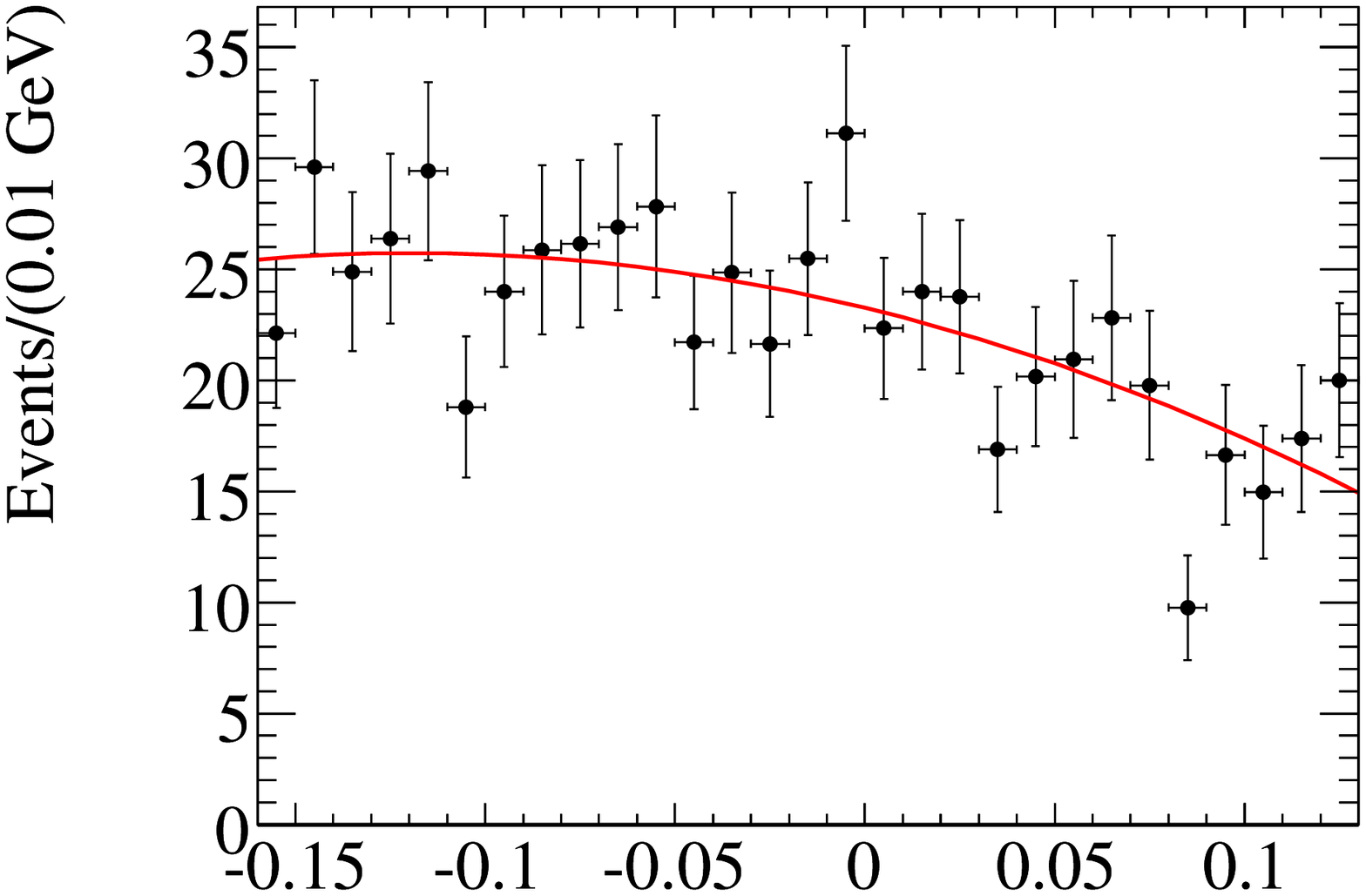}
\includegraphics[width=0.22\linewidth]{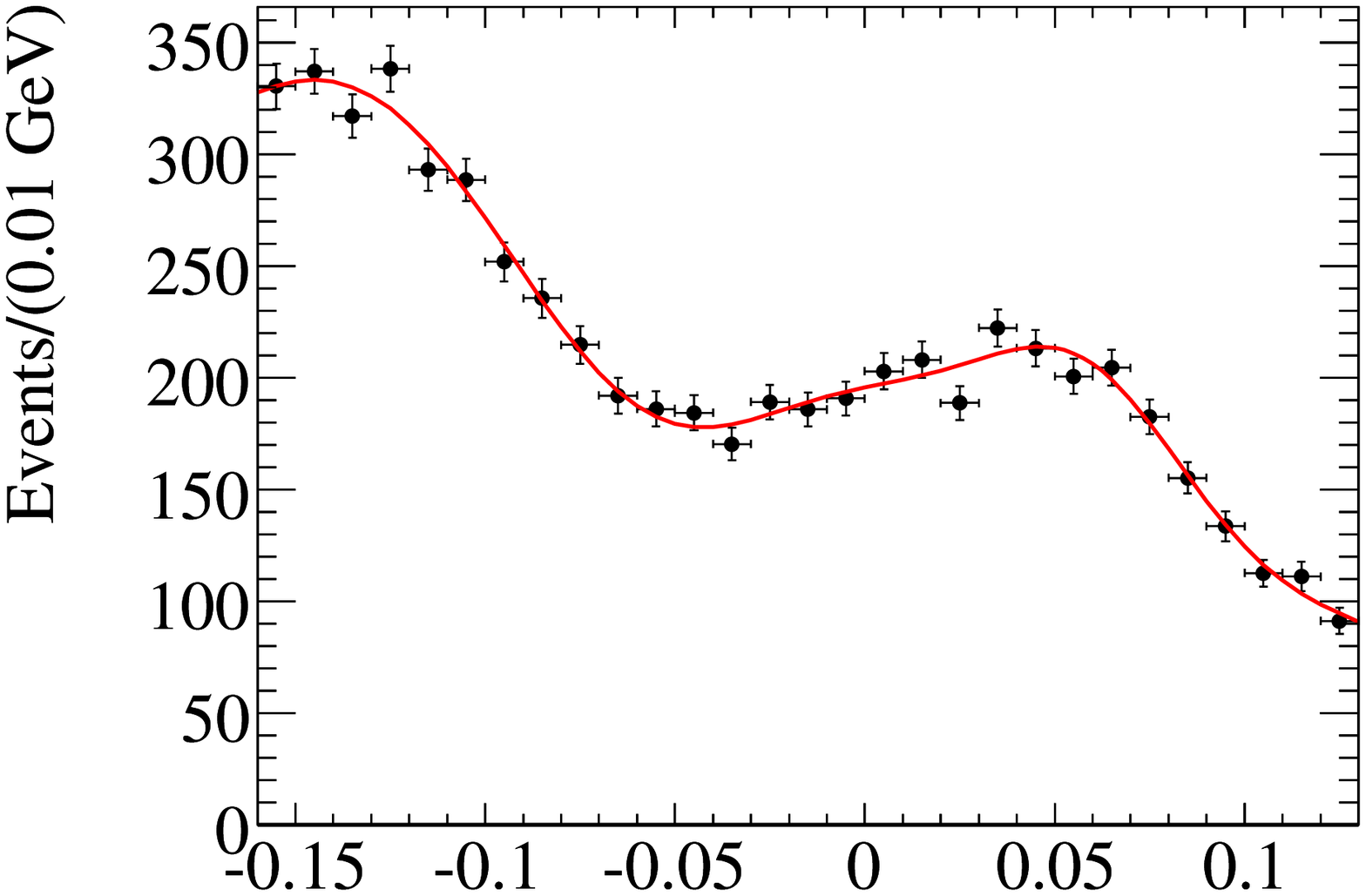}

$\DeltaE_K (\gev)$

\hspace{0.4cm}\includegraphics[width=0.23\linewidth]{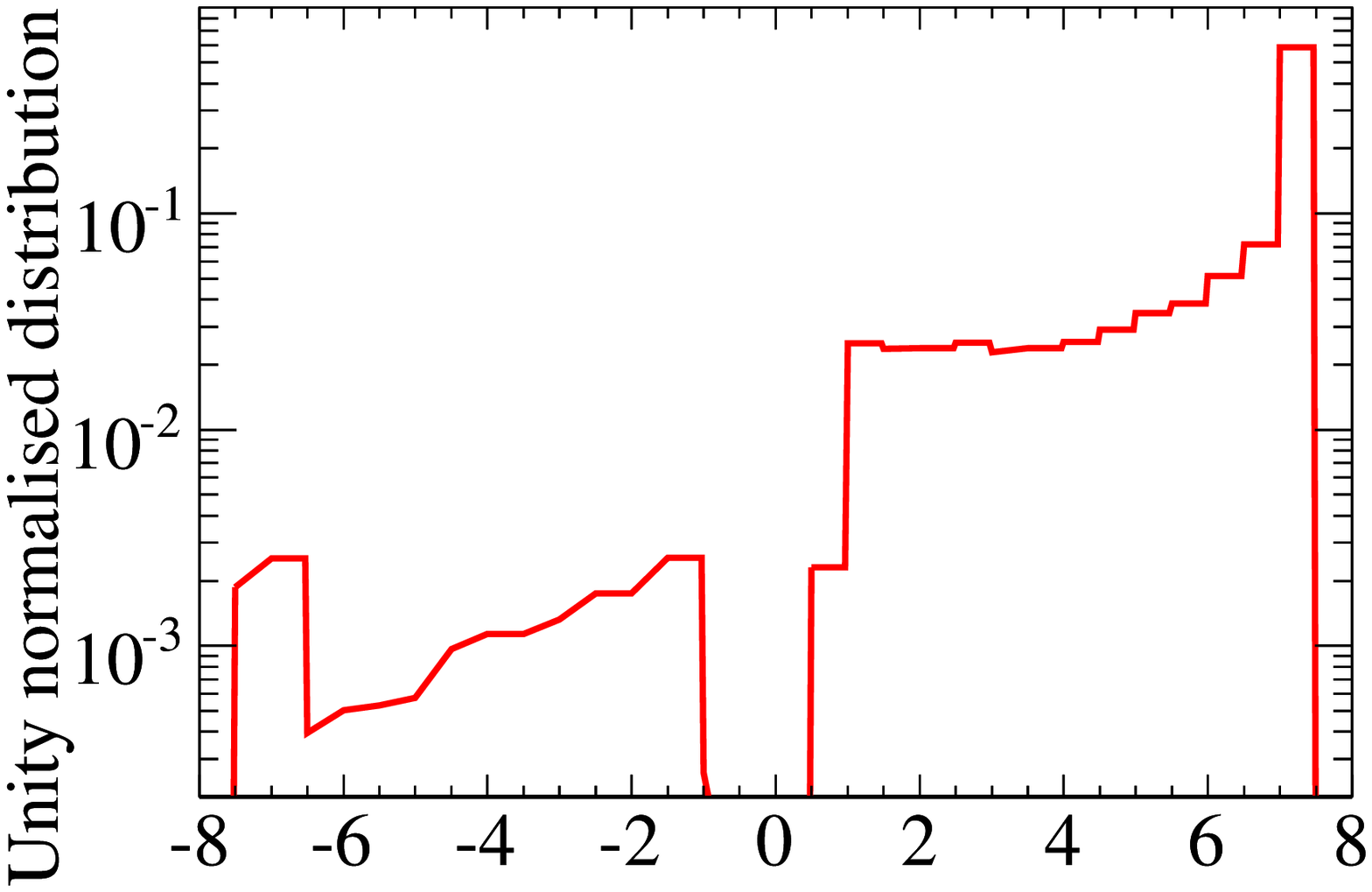}
\hspace{-0.2cm}\includegraphics[width=0.23\linewidth]{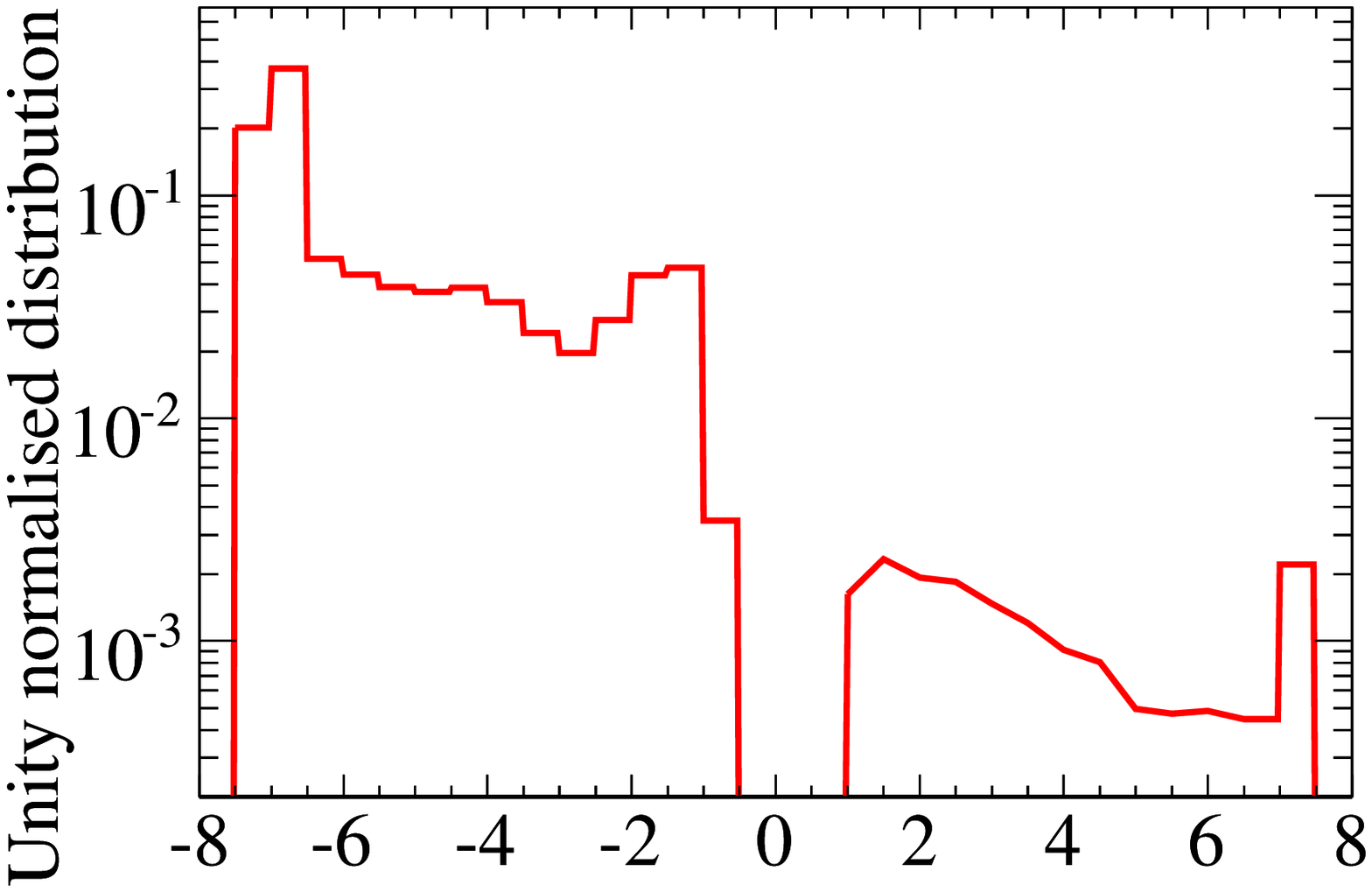}
\hspace{-0.2cm}\includegraphics[width=0.23\linewidth]{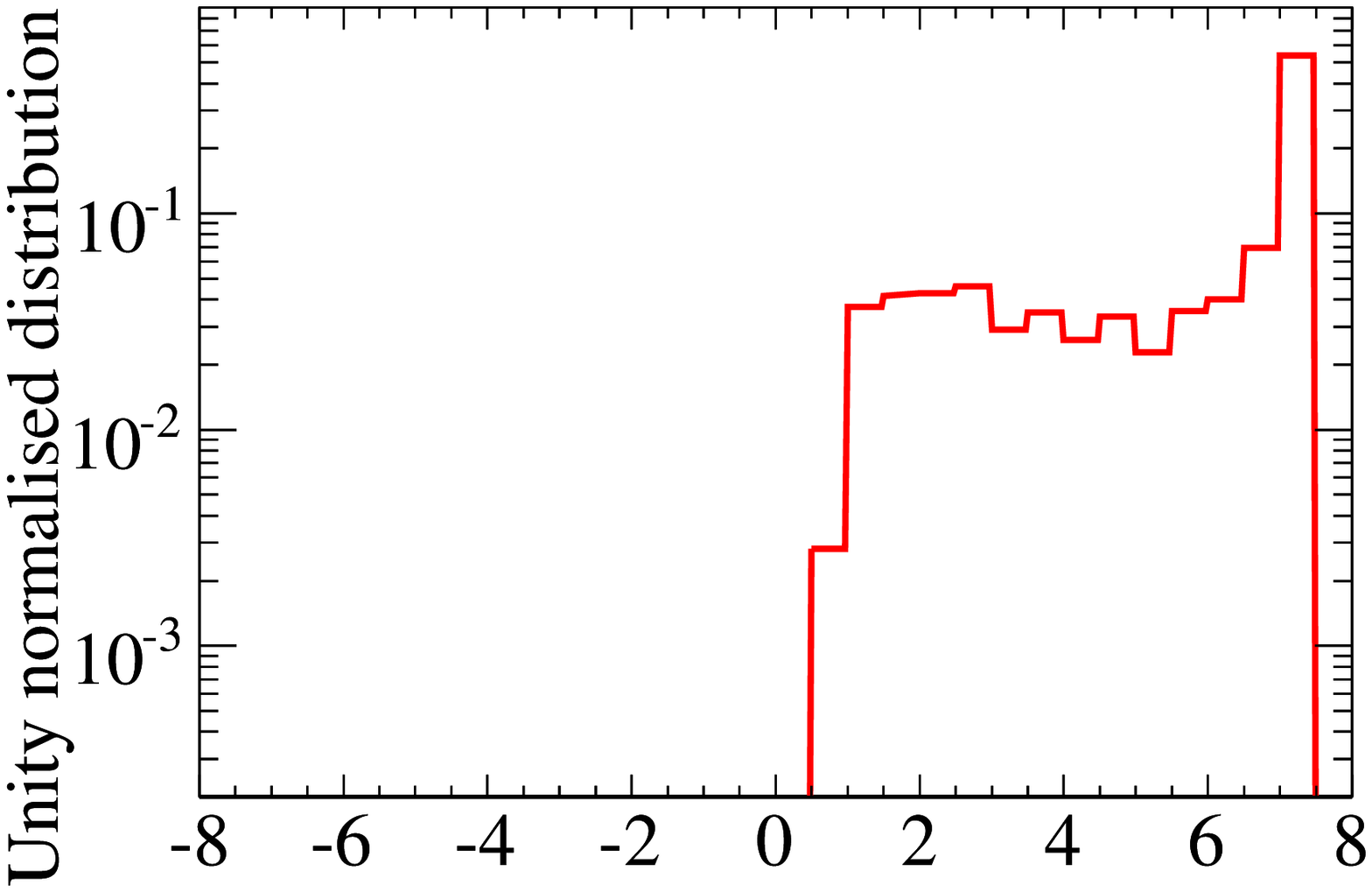}
\hspace{-0.2cm}\includegraphics[width=0.23\linewidth]{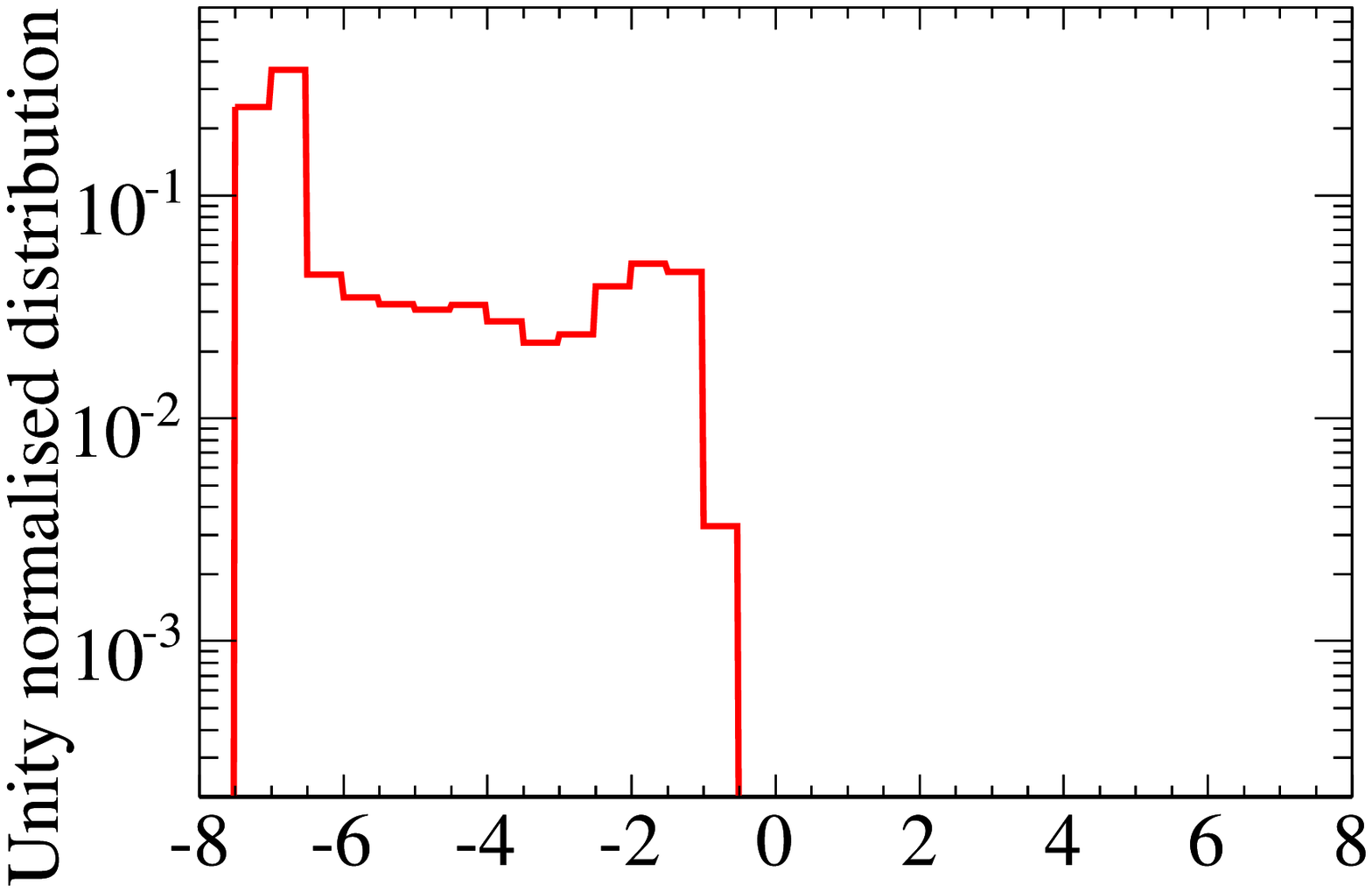}

$\mathcal{T_R}$

\caption{Distributions of $\DeltaE_K$  (upper plots) and the PID variable $\mathcal{T_R}$ (lower plots)
from MC simulations of the categories (from left to right)
$\Dstar K$,
$\Dstar\pi$, $B_K$ and $B_\pi$ (the latter two from  \BpBm,  \BzBzb and  \qqbar MC samples), for  the   mode $\Dstar \to D\piz, D \to\Kpm\pimp$.
In the upper plots, the dots represent the MC sample
spectrum, and the curves show the PDFs.  Note the vertical log scale in
the lower plots.
\label{fig:pdf:2420}}
\end{center}
\end{figure*}

The measurement is performed with an extended unbinned maximum likelihood function
\begin{equation}\label{eq:ELH}
\mathcal L=\frac{e^{-N'}(N')^N}{N!}\prod_{i=1}^N \mathcal P_i,
\end{equation}
where $N$ is the number of events in the sample to fit, $N'$ is
the expected number, and for event $i$
\begin{equation}
\label{eq:defPi}
\mathcal P_i= \frac{1}{N'} \sum_{j} N_{j}\mathcal P_i^{j},
\end{equation}
where $j=D^*K, D^*\pi, B_K, B_{\pi}$ is one of four events categories:
 signal kaon and pion, and background kaon and pion respectively,
where the background is a combination of continuum, $B^+B^-$ and
$B^0\overline B^0$ events.
The quantity $\mathcal P_i^{j}$ is the probability density function
(PDF) for event $i$ and category $j$, and $N_{j}$ is the number of
events in category $j$.

For the signal categories, the distance between the kaon and pion
$\Delta E_K$ peaks provides  powerful separation between pions and
kaons, in addition to PID.
For the background categories, we use mutually exclusive
likelihood-based pion and kaon selectors, that in particular contain
requirements of $\mathcal R>0.9$ (kaon) and $\mathcal R<0.1$ (pion)
respectively.
For consistency and symmetry reasons, the whole region $0.1 < \mathcal
R < 0.9$ is removed for all categories, including the signal
categories used in the fit.

The correlations between $\mathcal{T_R}$ and $\DeltaE_K$ are found to be small
(compatible with zero for the signal $K$ and for the background
categories, and with $-6\%$ for the $\pi$ signal category) therefore a
factorized approximation is used
\begin{equation}
\label{eq:defPdf2D}
\mathcal P_i^j(\Delta E_K,\mathcal{T_R})=\mathcal P_i^j(\Delta E_K)\mathcal P_i^j(\mathcal{T_R}).
\end{equation}
We have checked that no bias is introduced by this approximation by
simulating a large number of experiments in which the signal is taken from the large
statistics exclusive MC samples used for estimating these correlations.

The PDFs used in the fit are
determined from MC samples. The signal $\DeltaE_K$ PDFs are
parameterized with double Gaussian functions.
The background PDFs are mode-dependent functions chosen to best
 represent the MC background distributions: they include Gaussian,
 exponential, and third-order Chebyshev polynomial functions.
The complicated shape of the $\DeltaE_K$ distribution of the $B_\pi$ category arises from  the
contributions from several distinct components: at low $\Delta E_K$
values, $B^{\pm}\to D^*\rho^{\pm}$ decays dominate; in the signal
region, the background is mainly composed of $\gamma\leftrightarrow
\pi^0$ cross-feed and of $B^\pm \to \Dz \pipm$, the latter of which
dominates at high $\Delta E_K$ values.
The $\mathcal{T_R}$ PDFs are histograms, determined from MC
samples, with a binning $\Delta \mathcal{T_R} =0.5$ and, therefore, 28 bins.
MC-based studies have shown that the results of such fits do not
depend on the number of bins $n_b$  as long as $n_b>2$.

We correct for a small discrepancy in PID efficiencies between data and MC, using
high-statistics high-purity kaon and pion samples from inclusive
$\Dstarpm \to D \pipm$, $D\to \Kpm\pimp$ data.
The difference in track momentum spectra between these control samples
and the exclusive modes studied in the present analysis is accounted
for in the correction procedure.
This is achieved by weighting the control sample $\mathcal{T_R}$ PDF by
the ratio of the MC to control sample prompt track momentum distributions
for both cases of the prompt track being a kaon or a pion.
An example of the PDFs used for the channel $\Dstar \to D \piz$,
 $D \to\Kpm\pimp$ is shown in Fig. \ref{fig:pdf:2420}.

For signal events with a pion prompt track, for which  $\DeltaE_\pi$ (the subscript $\pi$ indicates that the pion hypothesis
has been assumed for the prompt track in the computation of \DeltaE) is close to zero, the relation
\begin{equation}
\label{eq:DeltaDeltaE}
\DeltaE_K - \DeltaE_\pi
 \approx \frac{1}{2p} \frac{E_{\FourS}}{m_{\FourS}}(m_K^2 - m_\pi^2),
\end{equation}
introduces a mild dependence of $\Delta E_K$ on the momentum $p$ of the
prompt track.
The parameters $E_{\FourS}$ and $m_{\FourS}$, $m_K$, $m_\pi$ denote the energy of
the $\epem$ system in the laboratory frame and the masses of the mesons, respectively.
Fits taking this dependence into account do not show any significant
improvement, nor degradation.

Fits performed on the \BpBm, \BzBzb and \qqbar background MC
 samples show no significant bias.
Similar fits with either pion or kaon signal events removed yield
 numbers of signal events compatible with zero for the removed category.
This indicates that the factorization approximation made for the
background PDF does not create any bias on the
number of fitted signal events.

Signal efficiencies are estimated from fits on high statistics exclusive MC
samples and summarized in Table \ref{tab:fit:eff}.
\begin{table}[h]
\caption{Selection efficiencies (in \%) for channels used in this analysis for
each decay mode of the $D$ (statistical uncertainties only).
\label{tab:fit:eff}}
\begin{center}
\begin{tabular}{lrrrr}
\hline
\hline\noalign{\vskip1pt}
 & $(D\pi^0)\Kpm$ & $(D\gamma)\Kpm$ & $(D\pi^0)\pipm$ & $(D\gamma)\pipm$ (in \%)\\
\hline\noalign{\vskip1pt}
$\Kpm\pimp$\hspace{1mm}& 21.0 \pom 0.1 \hspace{1mm}& 24.7 \pom 0.1 \hspace{1mm}& 22.2 \pom 0.1 \hspace{1mm}& 24.9 \pom 0.1 \\
$\pi\pi$   \hspace{1mm}& 14.6 \pom 0.1 \hspace{1mm}& 14.7 \pom 0.1 \hspace{1mm}& 14.8 \pom 0.1 \hspace{1mm}& 14.8 \pom 0.1 \\
$KK$       \hspace{1mm}& 20.4 \pom 0.1 \hspace{1mm}& 21.1 \pom 0.1 \hspace{1mm}& 20.5 \pom 0.1 \hspace{1mm}& 21.2 \pom 0.1 \\
$\KS\pi^0$ \hspace{1mm}&  8.9 \pom 0.1 \hspace{1mm}&  8.8 \pom 0.1 \hspace{1mm}&  8.9 \pom 0.1 \hspace{1mm}&  9.0 \pom 0.1 \\
$\KS\omega$\hspace{1mm}&  4.4 \pom 0.1 \hspace{1mm}&  4.2 \pom 0.1 \hspace{1mm}&  4.5 \pom 0.1 \hspace{1mm}&  4.3 \pom 0.1 \\
$\KS\phi$  \hspace{1mm}& 10.3 \pom 0.1 \hspace{1mm}& 13.5 \pom 0.1 \hspace{1mm}& 10.4 \pom 0.1 \hspace{1mm}& 13.7 \pom 0.1 \\
\hline
\hline
\end{tabular}
\end{center}
\end{table}
We perform separate fits for each \Dstar  decay mode, and subsequently perform a
weighted average to obtain our final results for $R^*_{\CP\pm}$ and  $A^*_{\CP\pm}$.
The free parameters of each fit are itemized here:
\begin{itemize}
\item
the charge-averaged $K/\pi$ ratio (one parameter, $R^*_{\pm}$ or $R^*$, whenever
relevant);
\item
the number of pion signal events (one parameter);
\item
the pion and kaon charge asymmetries $A_h^*$ of the signal  (two parameters);
\item
the number of pion and kaon background events, and charge asymmetries (four parameters);
\item
the position of the $\DeltaE_K$ peak of the pion events (one parameter).
\end{itemize}
In total there are nine free parameters for each \Dstar mode.

\section{Systematic uncertainties \label{sec:systematic}}

The systematic uncertainties are summarized in Table \ref{tab:systematics}.
\begin{table}[h]
\caption{Contributions to systematic uncertainties for each mode on the measurement
of the charge asymmetries $A_K^*$, and the ratio $R^*_{\CP}$ of \CP eigenmode to flavor
specific mode ($10^{-3}$).
See text for details.
\label{tab:systematics}}
\begin{center}
\begin{tabular}{rlrrrrrrr}
\hline
\hline
$\Dstar\to D \piz$
\\ \hline
 $ A_K^*$
\\
&            & $K\pi$& $\pi\pi$ & $KK$ & $\KS\pi^0$ & $\KS\omega$& $\KS\phi$ ($10^{-3}$) \\ \hline
& $\Delta E_K$ & 0 & 22 & 5 & 9 & 6 & 10 \\
& $\mathcal{T_R}$          & 1 & 12 & 2 & 18 & 22 & 38 \\
& $\BR (\Dstar \rho^-)$  & 3 & 4 & 0 & 1 & 2 & 59 \\
& $\BR (\Dstarm \pi^+)$  & 1 & 1 & 1 & 1 & 6 & 0 \\
& $\piz \leftrightarrow \g$  & 0 & 5 & 5 & 5 & 5  & 5 \\
& S-wave &  -- & --  & --  & --  & 33  &  2\\
\hline
& Total        & 3 & 26 & 7 & 21 & 41 & 71 \\
\hline \hline
 $R^*_{\CP}$
\\
&             & $K\pi$& $\pi\pi$ & $KK$ & $\KS\pi^0$ & $\KS\omega$& $\KS\phi$ ($10^{-3}$) \\ \hline
& $\Delta E_K$ & --   & 80 & 34 & 54 & 116 & 54 \\
& $\mathcal{T_R}$          & --   & 16 & 13 & 10 & 32 & 6 \\
& $\BR (\Dstar\rho^-)$  & -- & 51 & 22 & 14 & 5 & 131 \\
& $\BR(\Dstarm \pi^+)$  & -- & 3 & 5 & 1 & 21 & 4 \\
& S-wave &   -- & --  & --  & --  & 141 & 40 \\
\hline
& Total        & -- & 96 &  42 &  57 &  187 &  147  \\
\hline \hline
$\Dstar\to D \g$
\\ \hline
 $ A_K^*$
\\
&             & $K\pi$& $\pi\pi$ & $KK$ & $\KS\pi^0$ & $\KS\omega$& $\KS\phi$ ($10^{-3}$) \\ \hline
& $\Delta E_K$ & 1 & 39 & 8 & 19 & 66 & 52 \\
& $\mathcal{T_R}$          & 16 & 4 & 39 & 18 & 75 & 22 \\
& $\BR(\Dstar \rho^-)$  & 4 & 1 & 0 & 1 & 18 & 1 \\
& $\BR(\Dstarm \pi^+)$  & 4 & 3 & 4 & 0 & 18 & 3 \\
& $\piz \leftrightarrow \g$  & 0 & 10 & 10 & 10 & 10  & 10 \\
& S-wave &   -- & --  & --  & --  & 40  & 5  \\
\hline
& Total        & 17 &  40 &  41 &  28 &  111 &  58 \\
\hline \hline
 $R^*_{\CP}$
\\
&             & $K\pi$& $\pi\pi$ & $KK$ & $\KS\pi^0$ & $\KS\omega$& $\KS\phi$ ($10^{-3}$) \\ \hline
& $\Delta E_K$ & -- & 136 & 59  & 64 & 393 & 230\\
& $\mathcal{T_R}$          & -- & 17 & 34 & 4 & 78 & 42 \\
& $\BR(\Dstar \rho^-)$  & -- & 1 & 2 & 12 & 5 & 23 \\
& $\BR(\Dstarm \pi^+)$  & -- & 11 & 9 & 3 & 13 & 8 \\
& S-wave &   -- & --  & --  & --  & 192 &47  \\
\hline
& Total        & -- & 138 &  69 &  65 &  445 &  239 \\
\hline \hline
\end{tabular}
\end{center}
\end{table}

The contribution of the determination of the $\Delta E_K$ signal PDFs to
the systematic uncertainty is estimated by varying the
parameters that are fixed in the fit by one standard deviation
($\pm 1\sigma$).
The contribution of the $\Delta E_K$  PDFs of the $B_K$ category for the $\Dstarz\to
D\piz$ decays is determined similarly.

For the other $\Delta E_K$ background PDFs, this approach cannot be used, as the
correlations between parameters are not small. The contribution to the systematics due to the limited MC
statistics used to determine the parameters of the PDFs is
obtained in the following way. We determine two separate parameterizations of the PDFs on two halves
of the MC sample, and perform the fits with them.
We take half the difference between the obtained results as an
estimate of the  systematics.

The contribution of the determination of the $\mathcal{T_R}$ signal PDFs to the
systematic uncertainty is estimated by performing an additional fit
without the correction of the small discrepancy between data and MC described above.
The difference between the results of both fits is taken as an
estimate of the uncertainty.

The systematic uncertainty introduced by the limited knowledge of
$B^\pm\to D^*\rho^\pm$ and $B^0\to D^{*+}\pi^-$ branching fractions is
estimated from MC samples by performing a fit on a sample in which
the number of these events is varied by $\pm 1 \sigma$
\cite{ref:pdg06}.

Differences in the interactions of positively and negatively charged
kaons with the detector and the possible charge asymmetry of the
detector are studied using the exclusive MC samples.
Asymmetries of $(-1.0 \pom 0.2) \%$ and $(0.2 \pom 0.2) \%$ are observed for
kaon and pion modes, respectively, for the \CP modes.
A correction of +1\% is applied to the measured values of $A^*_{\CP}$.
The simulation of the detector charge asymmetry has been compared to
the actual value in the data in a previous analysis of \B decays to
$K\pi$
\cite{Aubert:2007mj}.
The possible discrepancy  has been found to be smaller than 1$\%$.

\begin{table*}[!Ht]
\caption{Summary of measurements of the charge asymmetries $A_K^*$;
 the \CP eigenmode to flavor specific mode ratios $R^*_{\CP}$, and
 the cartesian parameters $x^*$, for \Bpm decays to $\CP$ even and $\CP$ odd
 eigenmodes $\Dstar_{\CP} \Kpm$. The value of $A_K^*$ for the flavor-specific control
 mode with $D\to\Kpm\pimp$ is also given.}
\begin{center}
\begin{tabular}{lccc}
\hline
\hline
 & $A_K^*$ & $R^*_{\CP}$ & $x^*$ \\ \hline
Flavor specific\hspace{2mm}& $ -0.06 \pm 0.04 \pm 0.01$ \hspace{1mm}& -- & -- \\
$\CP+$ \hspace{1mm}& $ -0.11 \pm 0.09 \pm 0.01$ \hspace{1mm}& $1.31 \pm 0.13 \pm 0.04$ \hspace{1mm}& $0.11 \pm 0.06 \pm 0.02$ \\
$\CP-$ \hspace{1mm}& $ \phantom{-} 0.06 \pm 0.10 \pm 0.02$ \hspace{1mm}& $1.10 \pm 0.12 \pm 0.04$ \hspace{1mm}& $0.00 \pm 0.06 \pm 0.02$ \\
\hline
\hline
\end{tabular}
\end{center}
\label{tab:results}
\end{table*}

The $\piz\leftrightarrow \gamma$ cross-feed
can reduce the value of  $A^*_{\CP\pm}$ because for a given $D_{\CP}$ final
state, $D^*_{\CP}$ has the same \CP value as $D_{\CP}$ if decaying to
$D\piz$ and the opposite \CP value if decaying to $D\gamma$
\cite{Bondar:2004bi}. This ``\CP dilution'' is estimated from MC samples by
performing a fit in which the potential feed-across has been
completely removed. The effect is similar among  modes and is
accounted for by an uncertainty of $0.5\%$ for $D\piz$ modes and of
$1.0\%$ for $D\gamma$ modes.

In the case of $D$ decays to $\KS\phi$ and $\KS\omega$, the
\CP-violating charge asymmetry can be diluted by the presence of
decays to the same final state that may have a different \CP
composition ($\KS\Kp\Km$ and $\KS\pip\pim\piz$, respectively).
This S-wave effect is accounted for in a way similar to that used
in our previous study of the $D K$ modes \cite{Aubert:2005rw}.
It consists of applying a correction to the measured $A_{\CP\pm}^*$ and
$R_{\CP\pm}^*$ values using the \CP content information of $\KS\Kp\Km$ and $\KS\pip\pim\piz$ modes.
The uncertainty on the correcting factors is then propagated to the
correction formula and included as an additional systematic
uncertainty on $A_{\CP\pm}^*$ and $R_{\CP\pm}^*$.

The correlations between the different sources of systematic errors
are negligible and neglected when combining the two \CP-even or the
three \CP-odd modes.

No systematic error or correction is applied to account for selection
efficiency uncertainties as we do not measure branching fractions but
ratios of branching fractions in which they largely cancel.

For the branching fraction ratios $R_{\CP\pm}^*$, in addition to the
sources of systematic uncertainties listed in Table \ref{tab:systematics},
we associate one more uncertainty with
the assumption that $R_{\CP\pm}^*$ = $R_{\pm}^*$/$R^*$. This assumption holds only if the magnitude of the ratio $r_{\pi}^*$ between
the amplitudes of the $B^- \to \bar D^{*0} \pi^-$ and $B^-\to D^{*0}\pi^-$ processes
is neglected \cite{Gronau:2002mu}. The ratio $r_{\pi}^*$ is expected to be small:
$r_{\pi}^*\sim  r_B^* \frac{\lambda^2}{1-\lambda^2}$, where $\lambda\approx 0.22$ \cite{ref:pdg06} is the sine
of the Cabibbo angle. This introduces a relative uncertainty of
$\pm 2r_{\pi}^*\cos\delta_{\pi}^*\cos\gamma$ on $R_{\CP\pm}^*$, where $\delta_{\pi}^*$ is the
relative strong phase between $\mathcal A(B^- \to \bar D^{*0} \pi^-)$ and $\mathcal A(B^-\to D^{*0}\pi^-)$.
Since $|\cos\delta_{\pi}^*\cos\gamma|\leq 1$ and $r_{\pi}^*\leq 0.007$, we assign a relative uncertainty of $\pm 1.4\%$
to $R_{\CP\pm}^*$, which is completely anti-correlated between $R_{\CP +}^*$ and $R_{\CP -}^*$.

\begin{figure*}
\begin{center}
\begin{tabular}{cc}
 & \begin{tabular}{cc}
\large $\Dstar\to D\piz$\hspace{0.00cm} & \hspace{3.5cm}\large $\Dstar\to D\g$
\end{tabular}
\\
$K\pi$
&\begin{tabular}{cc}
\includegraphics[width=0.30\linewidth]{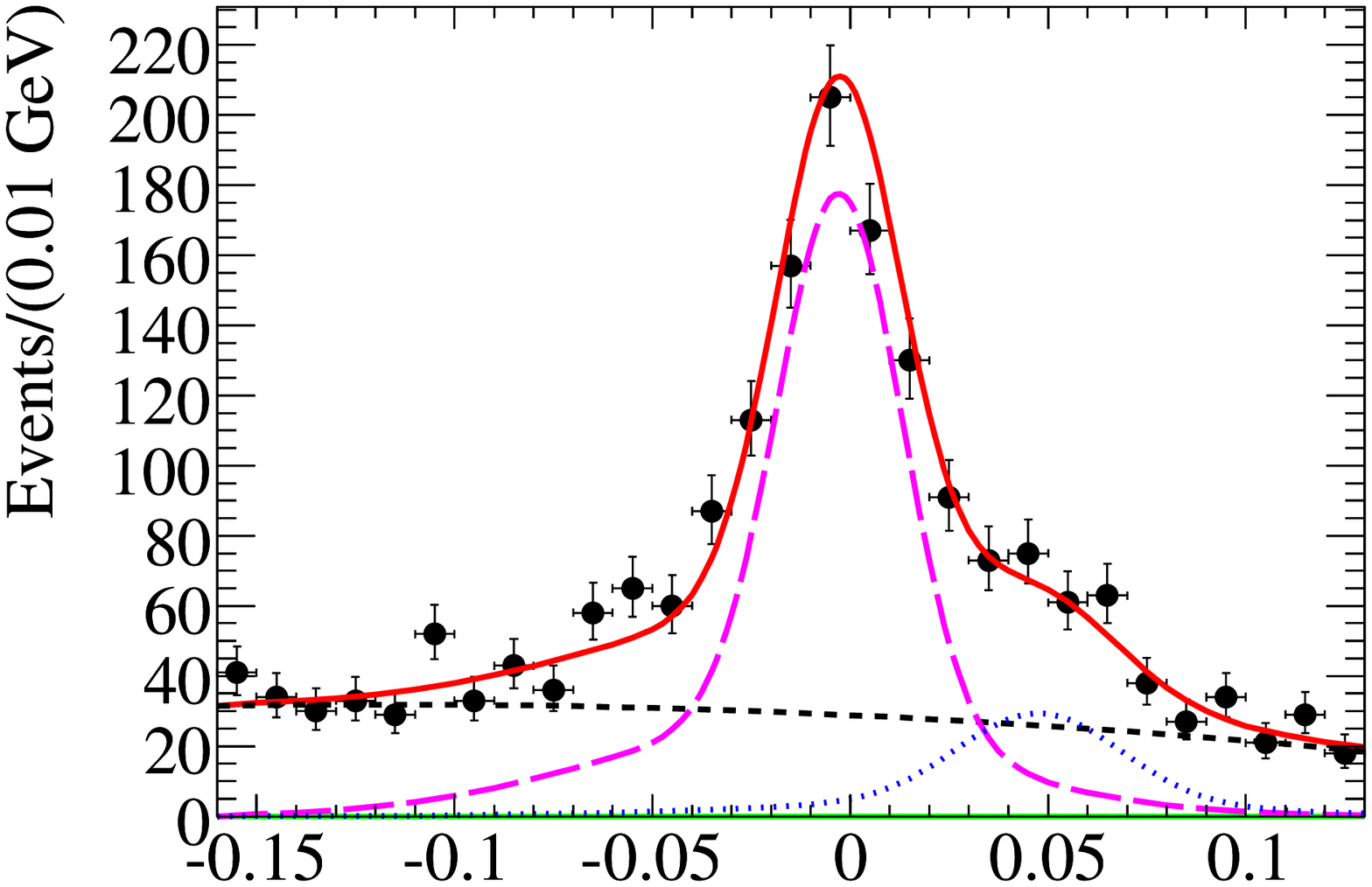}
& \includegraphics[width=0.30\linewidth]{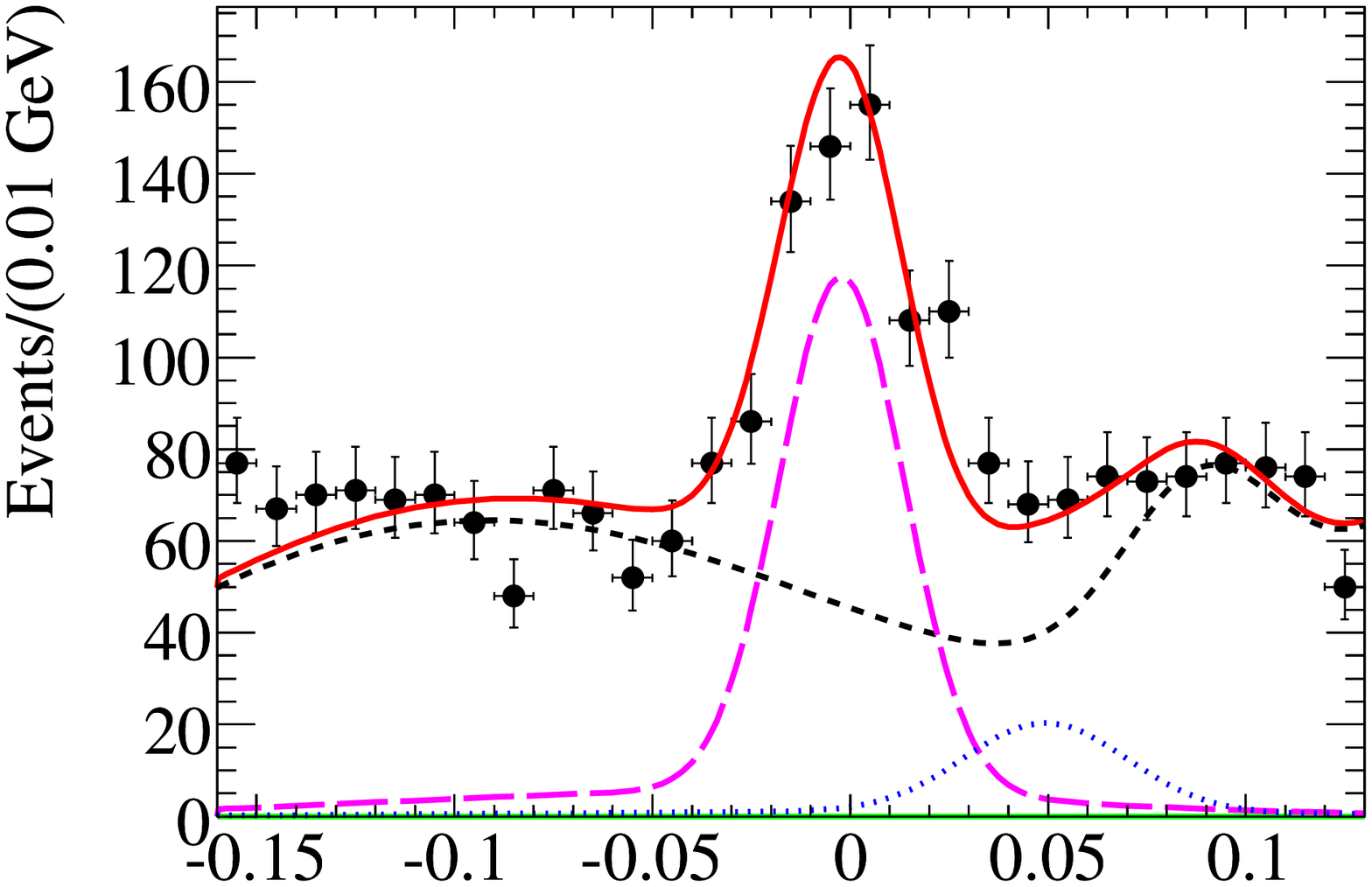}
\end{tabular}
\\
$\pi\pi$
& \begin{tabular}{cc}
\includegraphics[width=0.30\linewidth]{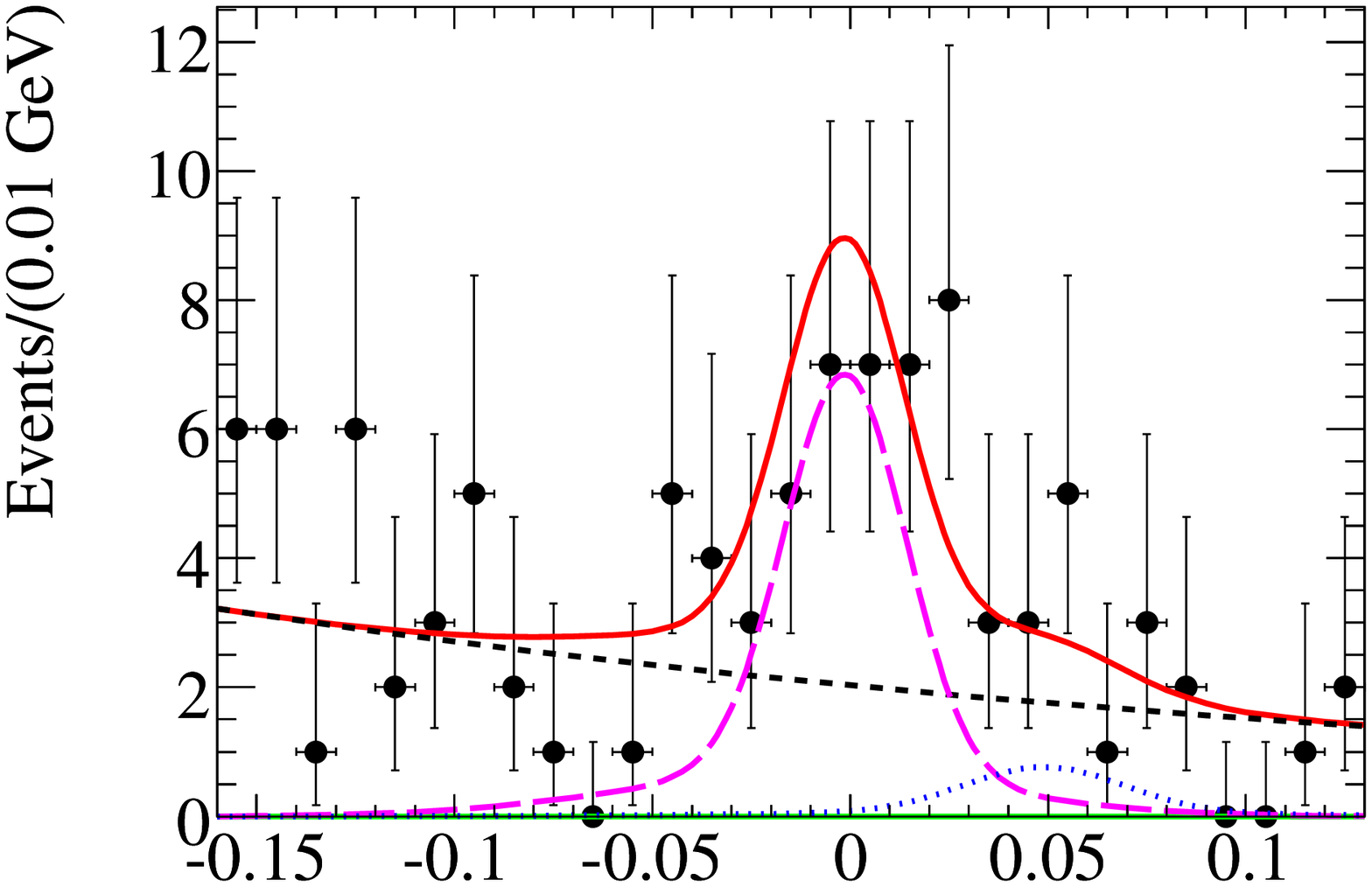}
& \includegraphics[width=0.30\linewidth]{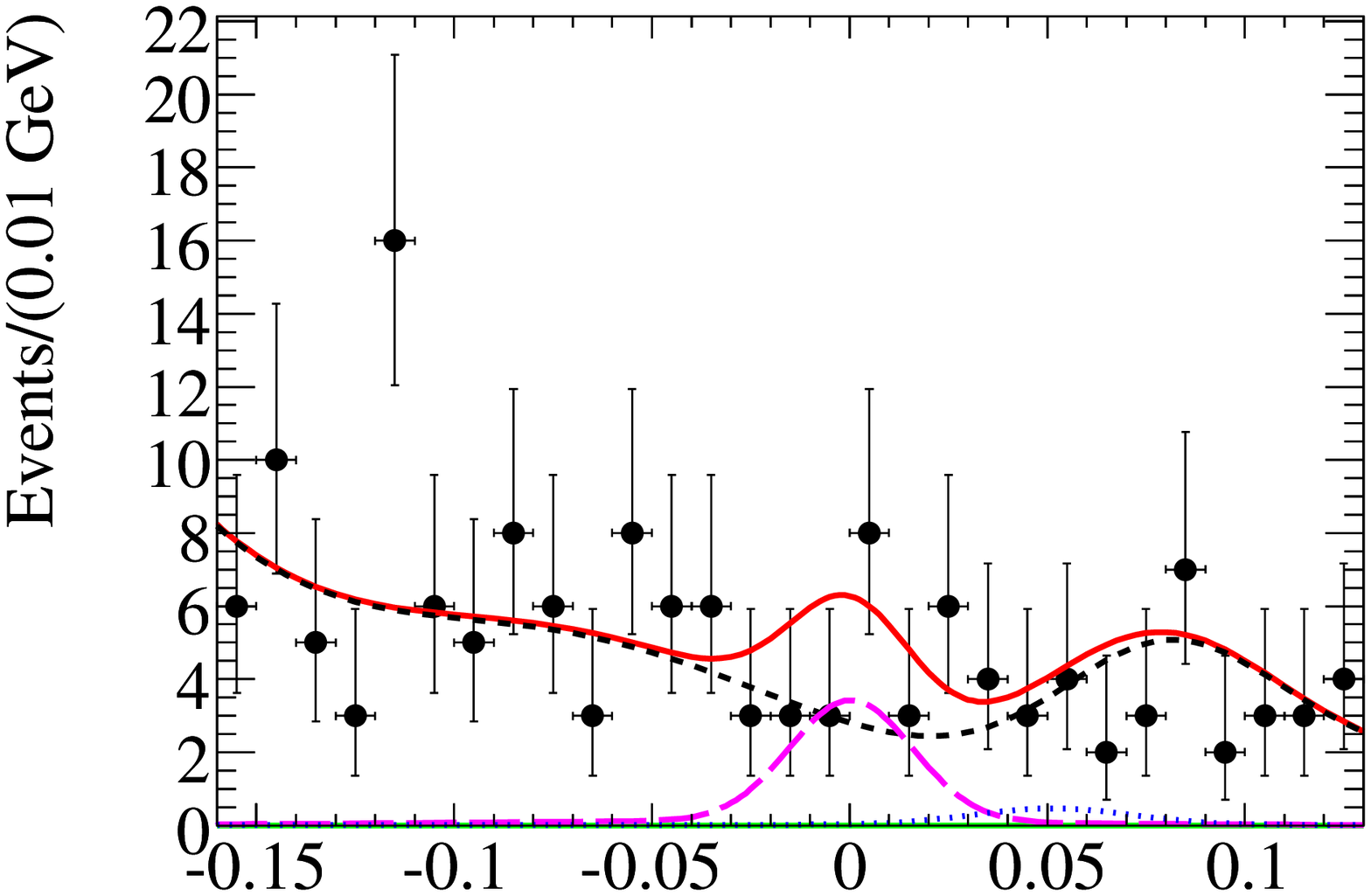}
\end{tabular}\\

$KK$
& \begin{tabular}{cc}
\includegraphics[width=0.30\linewidth]{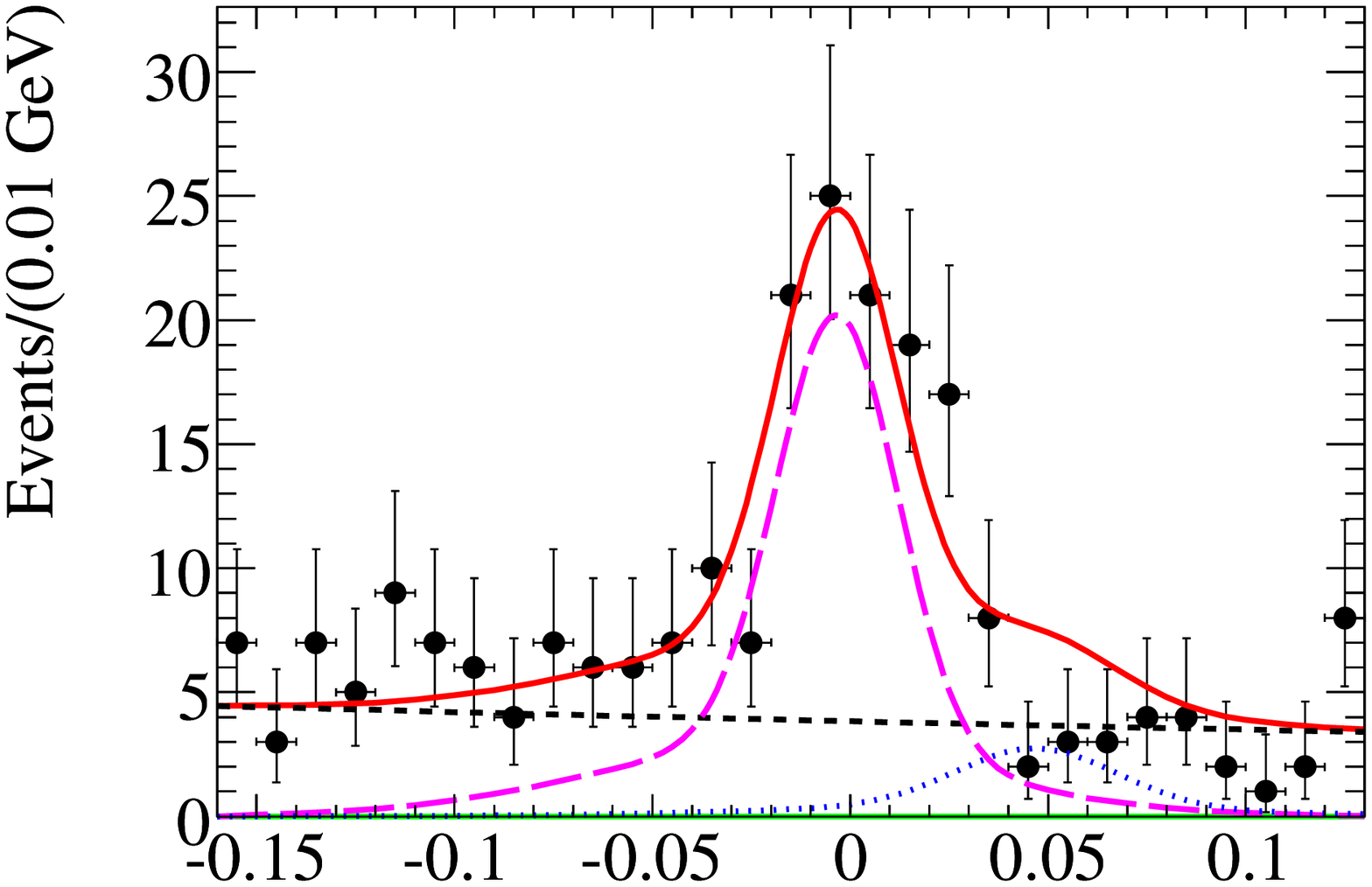}
& \includegraphics[width=0.30\linewidth]{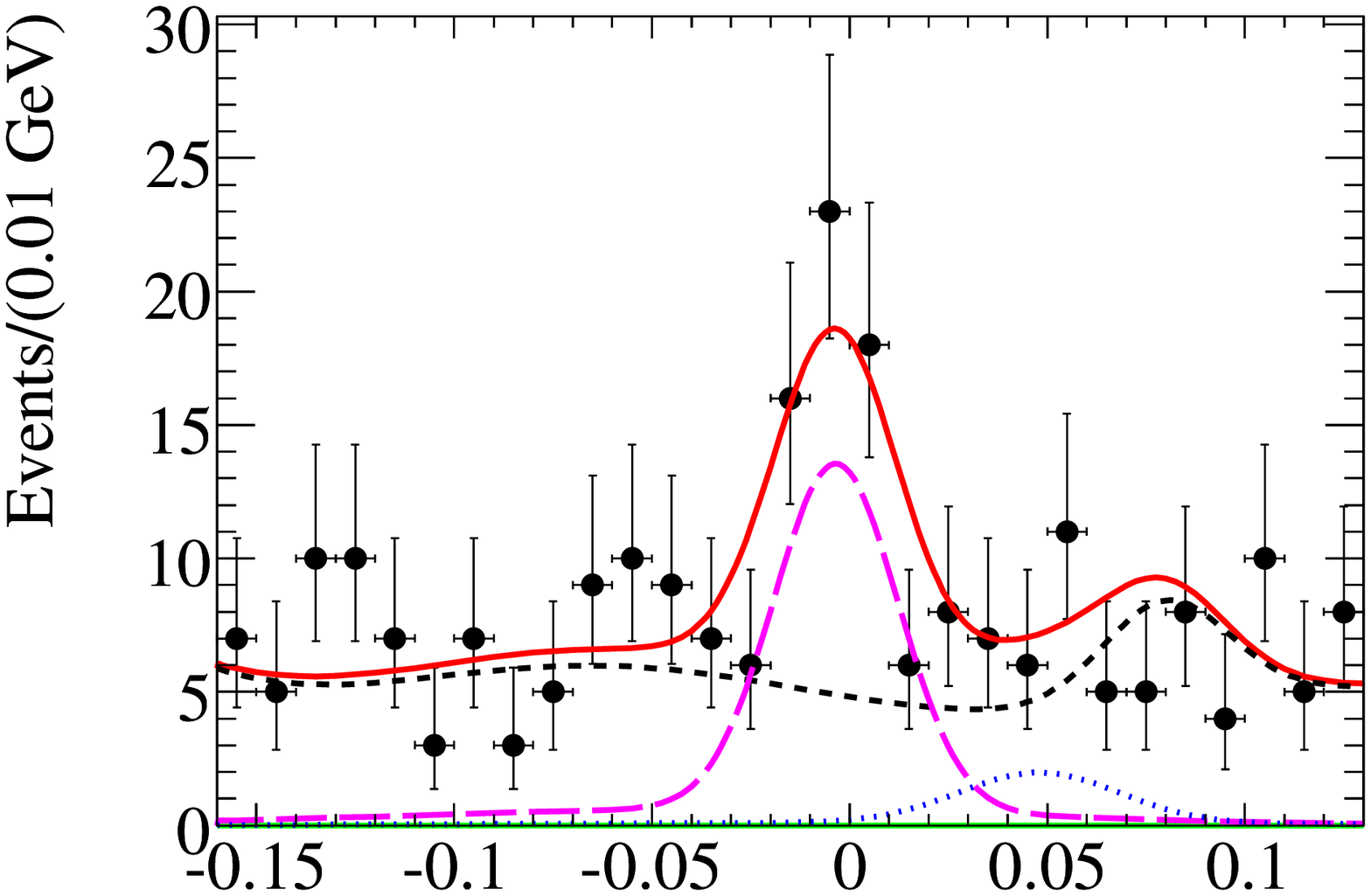}
\end{tabular}\\

$\KS \piz$
& \begin{tabular}{cc}
\includegraphics[width=0.30\linewidth]{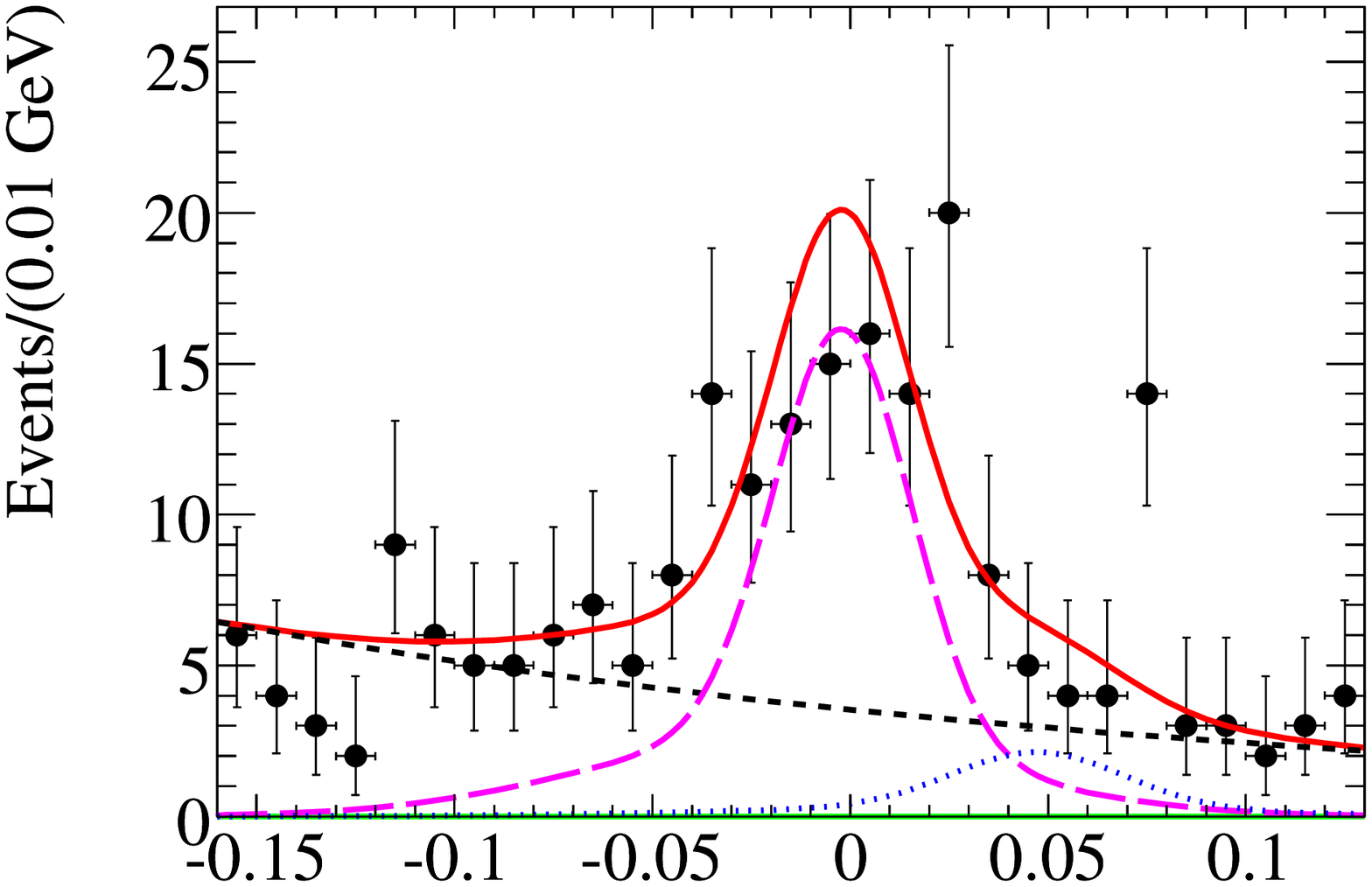}
& \includegraphics[width=0.30\linewidth]{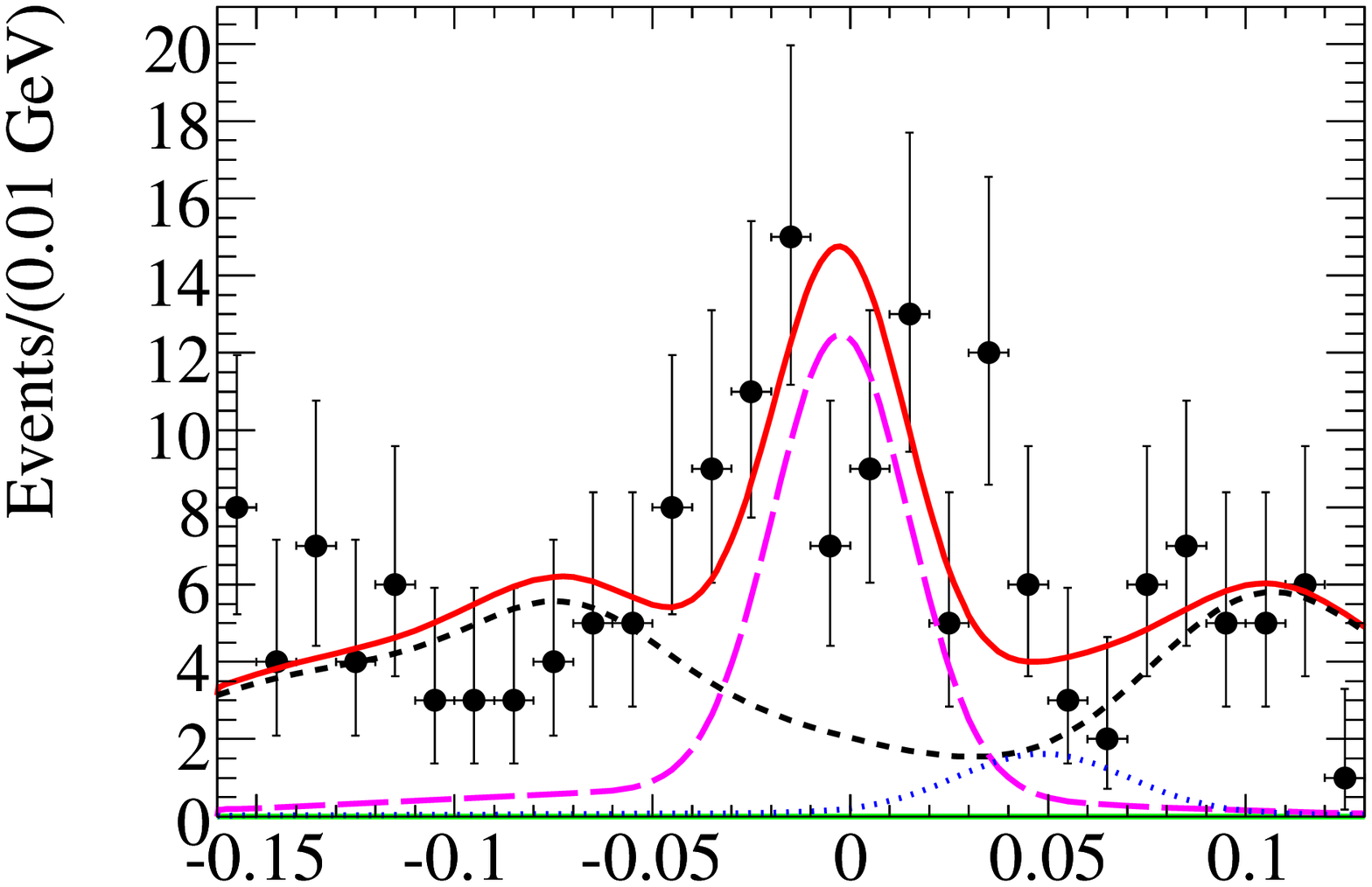}
\end{tabular}\\

$\KS \omega$
& \begin{tabular}{cc}
\includegraphics[width=0.30\linewidth]{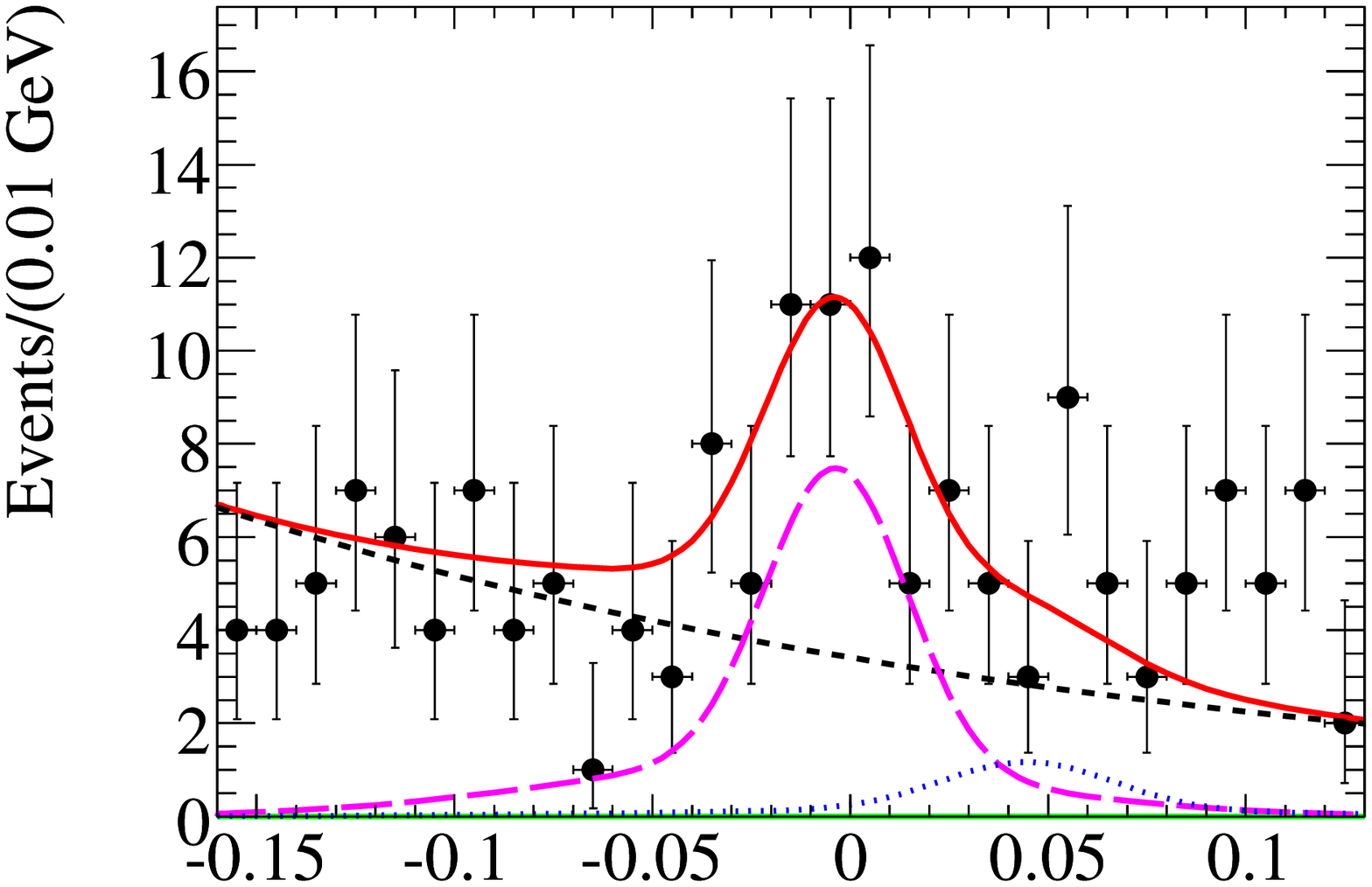}
& \includegraphics[width=0.30\linewidth]{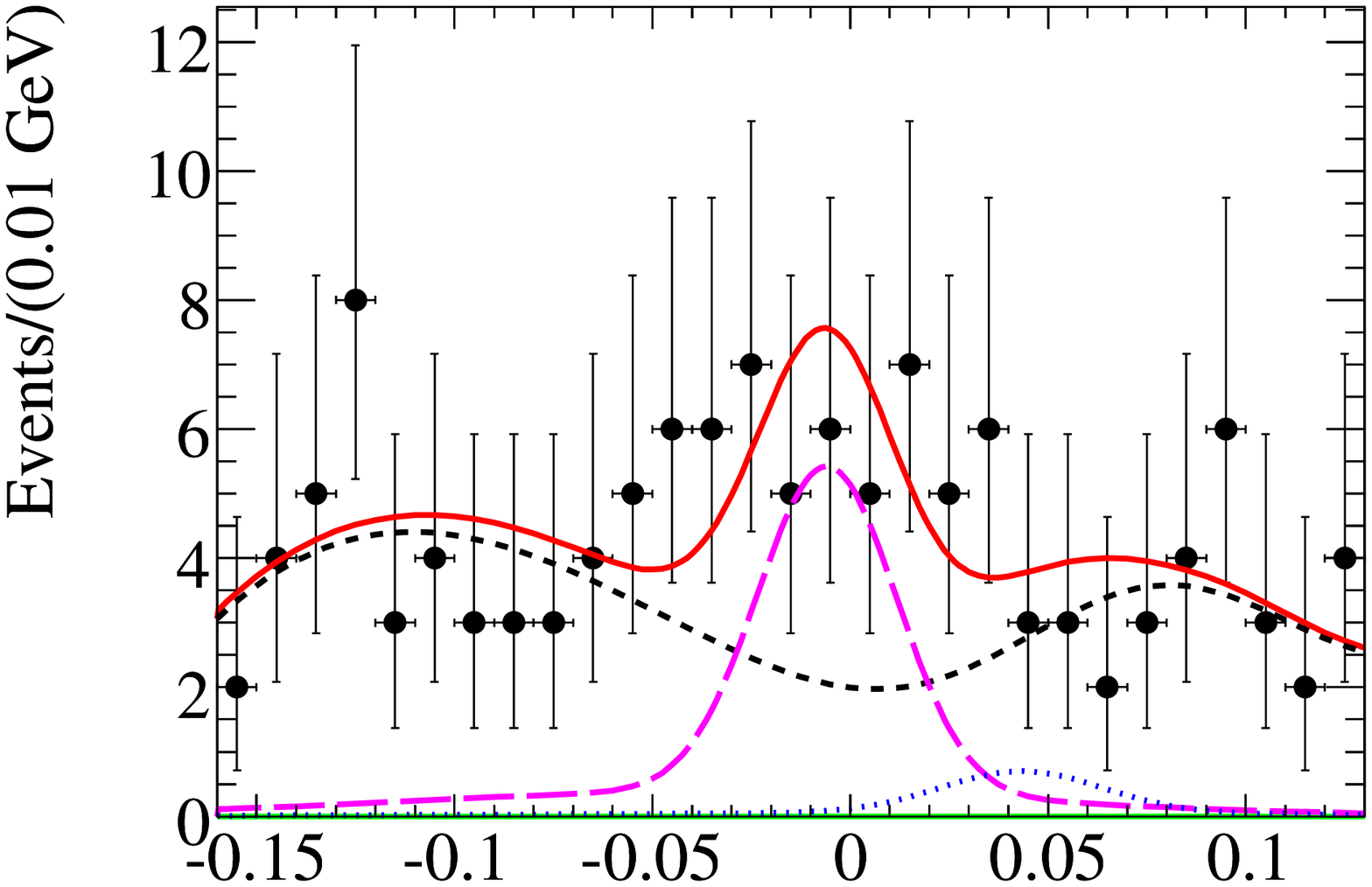}
\end{tabular}\\

$\KS \phi$
& \begin{tabular}{cc}
\includegraphics[width=0.30\linewidth]{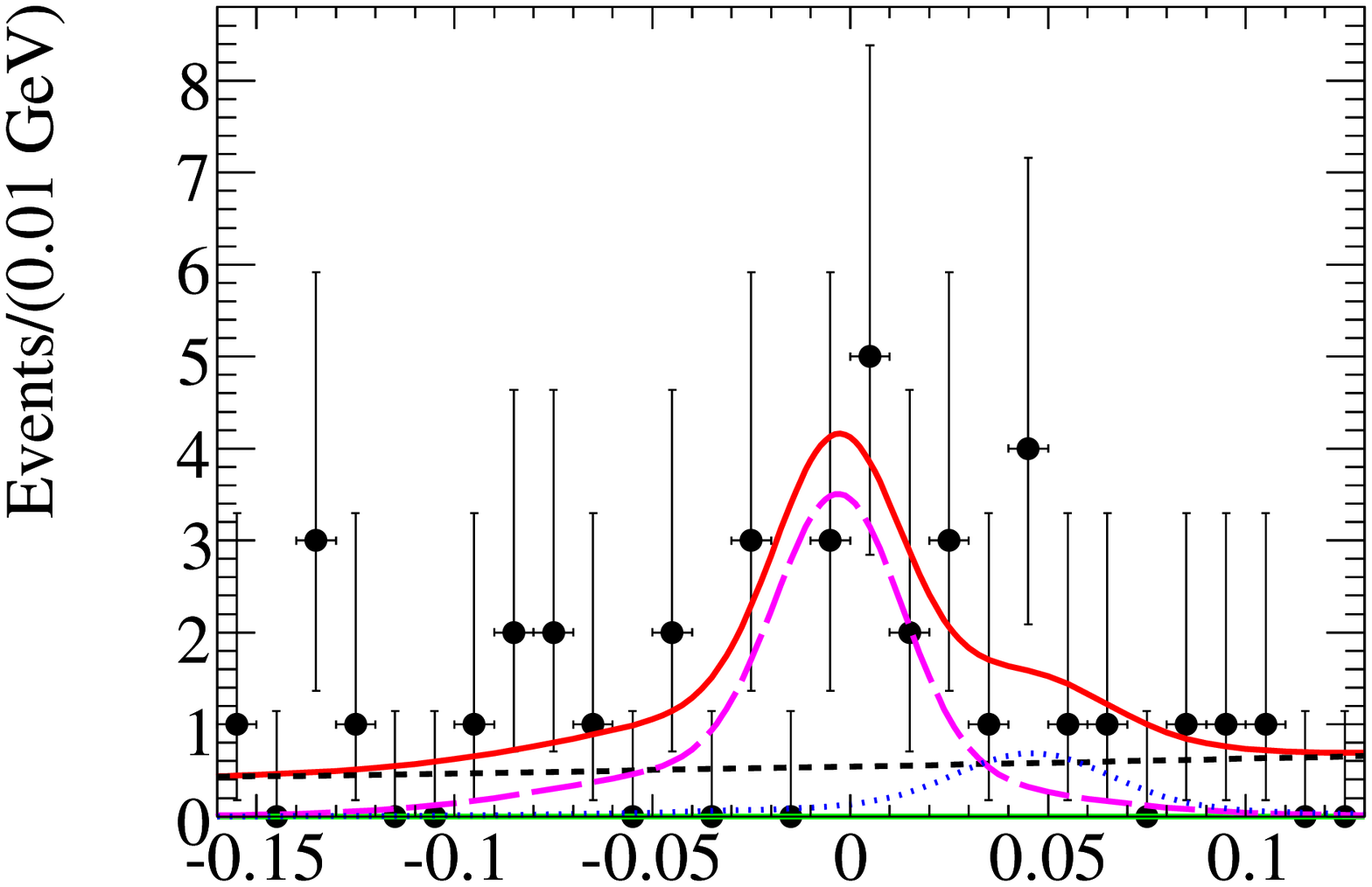}
& \includegraphics[width=0.30\linewidth]{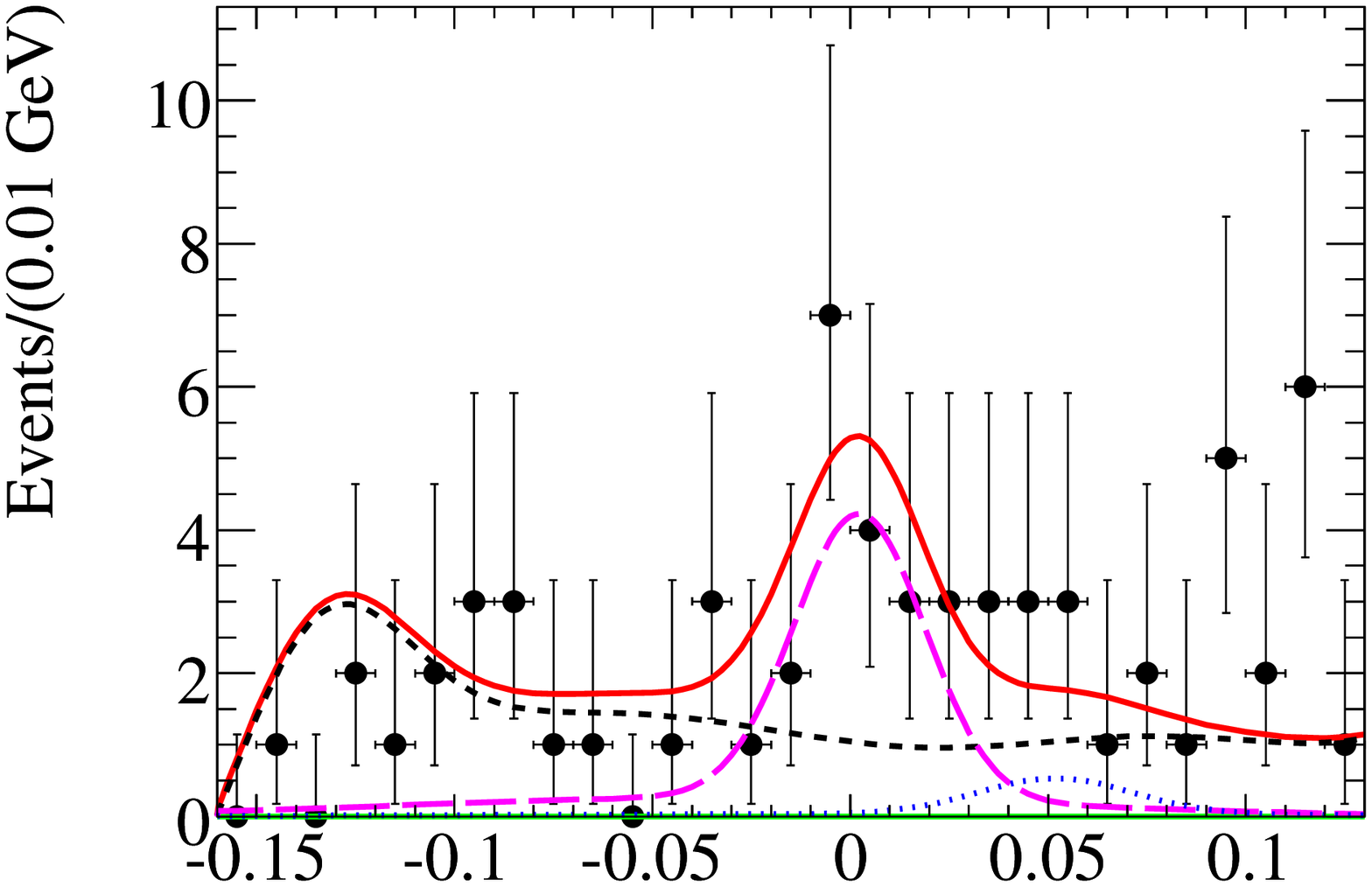}
\end{tabular}\\
\end{tabular}

\begin{tabular}{rr}
{ \hspace{2.0cm} $\DeltaE_K (\gev)$}& {\hspace{3.5cm}$\DeltaE_K (\gev)$}
\end{tabular}\\

\caption{
$\DeltaE_K$ distributions, with a cut on the PID variable $\mathcal{T_R}>0$ to enhance the kaon part of the sample.
Distributions are shown for each of the
decays modes: (left) $\Dstar\to D\piz$ and (right) $\Dstar\to D\gamma$, with the $D$ decay
modes indicated on the left of the figure.
Dots denote the distribution of the data.
Curves denote the PDFs for the various
 categories: signal $K$ (long-dashed curve);
signal $\pi$ (dotted curve); background $K$ (short-dashed curve);
background $\pi$ does not appear because the $\mathcal{T_R}>0$ cut completely removes this
category.
The thick curve denotes the total PDF.
\label{fig:data:bis}}
\end{center}
\end{figure*}

\section{Results \label{sec:results}}

We plot the $\DeltaE_K$ distributions in Fig. \ref{fig:data:bis} with
a kaon selection ($\mathcal{T_R}>0$) applied, and with the fitted PDFs overlaid.
The results are summarized in Table \ref{tab:results}, with the
observed numbers of charged-averaged events in Table \ref{tab:fit:nb:evts}.
Note that none of the corrections between data and MC that would be needed for
measurements of absolute branching fractions are used here, as we
are interested in ratios only.
We have checked that charge asymmetries of $B^\pm\to D^*\pi^\pm$ modes
are compatible with zero as expected:
$A^*_{\CP+,\pi} = 0.007 \pm 0.029 \pm 0.005$,
$A^*_{\CP-,\pi} = 0.032 \pm 0.027\pm 0.006$, and for flavor-specific modes
$A^*_\pi=-0.004\pm 0.010\pm0.001$.

\begin{table}[!htb]
\caption{Number of events measured in this analysis
 (statistical uncertainties only).
\label{tab:fit:nb:evts}}
\begin{center}
\begin{tabular}{lrrrr}
\hline
\hline\noalign{\vskip1pt}
 & $(D\pi^0)\Kpm$ & $(D\gamma)\Kpm$ & $(D\pi^0)\pipm$ & $(D\gamma)\pipm$ \\
\hline\noalign{\vskip1pt}
$\Kpm\pimp$ &$874\pm 44$ &$536\pm 36$ &$10729\pm 133$ & $7238\pm 119$\\ \hline
$\pi\pi$    &$31\pm 8$   &$15\pm 6$   &$262\pm 20$    &$170\pm 17$ \\ \hline
$KK$        &$101\pm 14$ &$62\pm 12$  &$987\pm 43$    &$709\pm 37$ \\ \hline
$\KS\piz$   &$86 \pm 14$ &$62\pm 11$  &$900\pm 38$    &$583 \pm 33$ \\ \hline
$\KS\omega$ &$43\pm 11$  &$29\pm 9$   &$419\pm 31$    &$250\pm 24$ \\ \hline
$\KS\phi$   &$19\pm 6$   & $21\pm 6$  &$262\pm 22$    &$180\pm 20$\\
\hline
\hline
\end{tabular}
\end{center}
\end{table}

We also obtain
$R^* = 0.0802 \pm 0.0031 \pm 0.0018$,
compatible with the theoretical prediction given in the introduction, and $(r^*_B)^2 =
0.20 \pm 0.09 \pm 0.03$.
We find
$\kappa = -0.08 \pm 0.16 \pm 0.02$
(defined in Eq.~\ref{eq:def:kappa}), consistent with zero as
expected.
We confirm the large value of $(r^*_B)^2$ that can be inferred from the previous measurements~\cite{Abe:2006hc} based on
the GLW method, with a precision improved by a factor
two.

Using the value of $\g=67.6\pm4.0$ obtained by a SM-based fit of the CKM matrix
\cite{CKMfitter} and the values of $r^*_B$ and $\delta^*_B$ from
Ref. \cite{Aubert:2006am},
we predict
 $A^*_{\CP+} = -0.18 \pm 0.10$,
 $R^*_{\CP+} = ~1.06 \pm 0.06$,
 $A^*_{\CP-} = 0.20 \pm 0.10$ and
 $R^*_{\CP-} = ~0.98 \pm 0.05$.
Our results are compatible with these predictions.

We also compute the cartesian coordinates with the channel $D \to \KS
\phi$ removed, so as to facilitate the comparison with results
obtained with the Dalitz method \cite{Aubert:2006am}:
\begin{eqnarray}
\label{eq:res:withoutKsphi}
x_{+}^* &=&  \phantom{-}0.09 \pm 0.07 \pm 0.02,  \\ \nonumber
x_{-}^* &=& -0.02 \pm 0.06 \pm 0.02,          \\ \nonumber
(r^*_B)^2 &=& \phantom{-}0.22 \pm 0.09 \pm 0.03.
\end{eqnarray}

\section{Summary \label{sec:summary}}

We have performed measurements of the \CP eigenmode to flavor specific
mode ratios and of the \CP-violating charge asymmetries of $\Bpm \to D^{*}\Kpm$ decays.
The ratios $R^*_{\CP}$ are found to be compatible with, and more precise than,
previous measurements.
Our results for $R^*_{\CP\pm}$ and $A^*_{\CP\pm}$ are at least a
factor of two more precise than previous measurements
\cite{Aubert:2004hu,Abe:2006hc}.
The precision of our results for $x^*_{\pm}$ is comparable to that
obtained from Dalitz plot analyses
\cite{Aubert:2006am,Collaboration:2008wy}.
No significant charge asymmetry is observed in the pion modes.
These results supersede our previous measurements \cite{Aubert:2004hu}.

\section{Acknowledgments \label{sec:acknowledgments}}

We are grateful for the 
extraordinary contributions of our \pep2\ colleagues in
achieving the excellent luminosity and machine conditions
that have made this work possible.
The success of this project also relies critically on the 
expertise and dedication of the computing organizations that 
support \babar.
The collaborating institutions wish to thank 
SLAC for its support and the kind hospitality extended to them. 
This work is supported by the
US Department of Energy
and National Science Foundation, the
Natural Sciences and Engineering Research Council (Canada),
the Commissariat \`a l'Energie Atomique and
Institut National de Physique Nucl\'eaire et de Physique des Particules
(France), the
Bundesministerium f\"ur Bildung und Forschung and
Deutsche Forschungsgemeinschaft
(Germany), the
Istituto Nazionale di Fisica Nucleare (Italy),
the Foundation for Fundamental Research on Matter (The Netherlands),
the Research Council of Norway, the
Ministry of Education and Science of the Russian Federation, 
Ministerio de Educaci\'on y Ciencia (Spain), and the
Science and Technology Facilities Council (United Kingdom).
Individuals have received support from 
the Marie-Curie IEF program (European Union) and
the A. P. Sloan Foundation.


\end{document}